\documentclass[
showpacs,preprintnumbers,
 amsmath,amssymb,
 aps,
 prd,
longbibliography,
floatfix,
lengthcheck, 
]{revtex4}

\usepackage{graphicx}
\usepackage{dcolumn}
\usepackage{bm}
\usepackage{listings}
\usepackage{natbib}
\usepackage[utf8]{inputenc}
\usepackage{hyperref}
\usepackage{color}
\usepackage{soul}

\begin{document}

\title{Nonstandard Neutrino Interactions in Supernovae}
\author{Charles J.~Stapleford}
\author{Daavid J.~V\"{a}\"{a}n\"{a}nen}
\author{James P.~Kneller}
\author{Gail C.~McLaughlin}
\affiliation{Department of Physics, North Carolina State University, Raleigh, NC 27695 USA}
\author{Brandon T. Shapiro}
\affiliation{Brandeis University, Waltham, MA 02453 USA}

\date{\today}

\begin{abstract}
Nonstandard Interactions (NSI) of neutrinos with matter can significantly alter neutrino flavor evolution in supernovae with the potential to impact explosion dynamics, nucleosynthesis and the neutrinos signal. 
In this paper we explore, both numerically and analytically, the landscape of neutrino flavor transformation effects in supernovae due to NSI and find new, heretofore unseen transformation processes can occur. These new transformations can take place with NSI strengths well below current experimental limits. Within a broad swath of NSI parameter space we observe Symmetric and Standard Matter-Neutrino Resonances (MNRs) for supernovae neutrinos, a transformation effect previously only seen in compact object merger scenarios; in another region of the parameter space we find the NSI can induce neutrino collective effects in scenarios where none would appear with only the standard case of neutrino oscillation physics; and in a third region the NSI can lead to the disappearance of the high (H) density Mikheyev-Smirnov-Wolfenstein (MSW) resonance. Using a variety of analytical tools we are able to describe quantitatively the numerical results allowing us to partition the NSI parameter according to the transformation processes observed. Our results indicate nonstandard interactions of supernova neutrinos provide a sensitive probe of Beyond the Standard Model physics complementary to present and future terrestrial experiments.  
\end{abstract}
\medskip

\maketitle

\section{Introduction}
\label{sec:intro}

The pursuit of Beyond the Standard Model (BSM) physics is a major goal of current nuclear and high energy physics research. Investigations of such phenomena as dark matter, the matter-antimatter asymmetry and neutrino mass and mixing are presently being explored. A lucrative source of information about BSM physics has been the neutrino which has yielded significant discoveries in the form of neutrino mass and mixing. Ongoing and future study of neutrinos may yield evidence for proposed BSM physics such as new interactions of active flavors, the origin and the nature of neutrino mass, additional flavors, CP violation and more. Much of this search will be conducted using experiments here on Earth, see, for example, \cite{1999PhRvL..82.3202G,2004PhRvD..70k1301F,PhysRevD.88.073012,2013JHEP...06..026E,2014PhRvL.113n1802A,2010JHEP...06..068A,2009PhLB..671...77W,2015arXiv151105562D,2015NuPhB.893..376M,2008PhLB..668..197E,2014PhLB..739..357C}. However, much also can be learned from studying the effect of BSM physics upon neutrinos in astrophysical environments for the simple reason that in the cores of supernovae, in the early universe, and in the mergers of compact objects, the densities, temperatures, magnetic fields, etc.\ can be so high that the neutrino is no longer an ephemeral component of the system but rather becomes an important mechanism for transporting energy and momentum as well as playing the familiar role of modifying the electron fraction. In essence, supernovae, compact object mergers, and the early universe constitute nature's ultimate neutrino experiment: if we change the properties of the neutrino, there can be major consequences for dynamics of the system, the nucleosynthesis, and significant modifications to any signal we might detect. 

There are two significant benefits to studying neutrinos emitted from both core-collapse supernova and compact object mergers: firstly the neutrino flavor evolution is non-linear due to neutrino collective effects \cite{Duan:2006an,Duan:2006jv} allowing seemingly small perturbations to become amplified, and secondly,  core-collapse supernova and mergers produce so many neutrinos a Galactic supernova or merger may produce sufficient events in current and future generation neutrino detectors to reveal the BSM physics. The effect of sterile neutrinos in supernovae has been considered on many occasions \cite{1997PhRvD..56.1704N,1999PhRvC..59.2873M,2001NuPhB.599....3P,2006PhRvD..73i3007B,2012JCAP...01..013T,2014PhRvD..90j3007W,2014PhRvD..89f1303W,2014PhRvD..90c3013E} and neutrino magnetic moments were studied by \cite{1988PhRvL..61...27B,1999PhLB..470..157M,1999APh....11..317N,2007JCAP...09..016B,2012JCAP...10..027D}. Various nonstandard interactions of neutrinos with matter have been considered \cite{1987PhLB..199..432V,1996PhRvD..54.4356N,1996NuPhB.482..481N,1998PhRvD..58a3012M,2002PhRvD..66a3009F} in particular we mention those by Esteban-Pretel, Tom{\`a}s, and Valle \cite{PhysRevD.76.053001} looking at the modification of the MSW effect in supernovae \cite{PhysRevD.17.2369,Mikheyev:1985aa,Mikheyev:1986tj}, and then again by Blennow, Mirizzi and Serpico \cite{2008PhRvD..78k3004B} and Esteban-Pretel, Tom{\`a}s, and Valle \cite{2010PhRvD..81f3003E} where the neutrino-neutrino interactions were included. These studies showed the effect of NSI is to introduce new contributions to the matter potential seen by the neutrinos with strengths (relative to the standard neutrino oscillation contribution to the potential) parameterized by a set of matrices. In general these matrices of NSI strengths can be different for each constituent of the matter and can contain \lq\lq off-diagonal" contributions also known as flavor changing neutral currents. It was shown that the new NSI contributions to the matter potential can lead to new MSW resonances close to the proto neutron star that were named as inner (I) resonances \cite{PhysRevD.76.053001}. 

As we shall explore in this paper, the presence of NSI completely changes the flavor oscillations.While some of these effects are already known, in this work we consider a larger NSI parameter space than was considered previously and find in the unexplored regions a new set of oscillation phenomena not previously reported. In one part of the parameter space we observe Matter-Neutrino Resonance (MNR) transitions. In another we find neutrino-neutrino collective effects in scenarios when none are observed without NSI, and in a third region we find the high (H) density MSW resonance can disappear. 

MNR transitions are a recently discovered phenomenon that can be explained as an active cancellation of the neutrino-neutrino and the background matter potential~\cite{2012PhRvD..86h5015M}. Two types of MNR have been observed. In a Standard MNR both neutrinos and antineutrinos start converting but end up in opposite flavor configurations while in a Symmetric MNR both neutrinos and antineutrinos fully convert to other flavors \cite{Malkus:2014iqa,Wu:2015fga,Vaananen:2015hfa}. The implications of MNR on nucleosynthesis in neutron star merger scenarios were further investigated in Refs.~\cite{Malkus:2015mda,Frensel:2016fge}. The possible effects of MNR have also been explored in the early universe~\cite{Johns:2016enc}.
While the MNR has been seen with standard neutrino oscillation physics alone, neither a Standard nor Symmetric MNR can occur in a supernova with only the standard neutrino physics if the neutrino emission is spherically symmetric. Beyond the neutrinosphere there are no locations where the conditions necessary for the MNR prevail. But as we shall show, both Symmetric and Standard MNRs can occur in supernovae in the presence of NSI. In particular, the NSI induced Standard MNR appears to be a very robust phenomenon and appears over a large part of the overall parameter space explored. 

Bipolar type flavor conversions, also known as nutations, were first observed by Duan \emph{et al.}\cite{Duan:2006jv,Duan:2006an} and are still an active area of research: see \cite{Duan:2009cd,Duan:2010bg} for reviews. In a bipolar/nutation type flavor transformation a large fraction of the whole ensemble of neutrinos can experience coherent oscillations with respect to each other converting a neutrino of one initial flavor into another. Bipolar/nutation type transitions are not adiabatic transitions in the sense that neutrinos do not follow the instantaneous eigenstates of the Hamiltonian. While these type of flavor transitions also occur with standard neutrino oscillation physics, we shall show they are affected by NSI.

The H MSW resonance was first introduced by Dighe \& Smirnov \cite{2000PhRvD..62c3007D} who noted that for three flavor neutrino mixing there were two distinct resonances, one at low (L) density and the other at high density. The width and eigenvalue splitting at the low density resonance are controlled by the mixing parameters $\delta m_{21}^2$ and $\theta_{12}$ while the H resonance was set by the mixing parameters $\delta m_{31}^2$ and $\theta_{13}$. Due to the uncertainty in the sign of the mass splitting, the H resonance would appear in the neutrinos for a normal hierarchy when $\delta m_{31}^2 >0$, and in the antineutrinos if $\delta m_{31}^2<0$. Like bipolar/nutation type transitions, the H resonance occurs with standard neutrino oscillation physics. We shall show it too can be affected by NSI and, in certain regions of the parameter space, it can disappear. 

With our goal defined, this paper is organized as follows. We begin with a description of a model in section \S\ref{sec:setup} and describe the origin of the effects we find in Section \S\ref{sec:Matter} which will allow us to diagnose which type of transition we find when we show results some specific selected combinations of NSI parameters in Section \S\ref{sec:effects}. In section \S\ref{sec:partition} we present our partitioning of the NSI parameter space and use our analytical tools to explain why the effects we found in each region were seen. In our Discussions and Conclusions \S\ref{sec:conc} we indicate the possible implications of NSI for both the neutrino signal and the dynamics of the explosion that we shall pursue in future studies.

\section{Model Details}
\label{sec:setup}

Our intention in this paper is to illustrate the possible effects of NSI in supernovae and highlight novel aspects that have not been studied in earlier literature. In order to effectively demonstrate these effects, we adopt a simplified model of supernova neutrinos using parameterized density and electron fractions, two neutrino flavors (electron and other-than-electron type which we shall denote these by `e' and `x' hereafter), a single energy and we make use of the single angle approximation for the neutrino self interaction. These simplifications remove the complicating factors of a more complete calculation but retain the essential physics. Zhu, Perego, and McLaughlin \cite{2016arXiv160704671Z} have compared the results from two and three flavor calculations of the Matter Neutrino Resonance above accretion disks and found they give very similar results if the gradients of the potentials are not large. For ease of computation we will utilize flux normalized density matrix formalism for both neutrinos and antineutrinos. We normalize the density matrices with respect to the initial electron neutrino flux and the initial electron antineutrino flux so that our initial conditions are 
\begin{equation}
	\rho(0) =  \frac{1}{1+\beta} \left(
		\begin{array}{cc}
		1 & 0 \\ 
		0 & \beta 
		\end{array}
	\right) \ , \quad\quad
	\bar{\rho}(0) = \frac{1}{1+\bar{\beta}}\left(
		\begin{array}{cc}
		1 & 0 \\ 
		0 & \bar{\beta} 
		\end{array}
	\right) \ , 
    \label{eq:InitialCondition}
\end{equation}
for neutrinos and antineutrinos respectively,
where $\beta$ represents the initial asymmetry between electron and x-type neutrinos, and $\bar{\beta}$ the asymmetry between electron and x-type antineutrinos.
The ratio of electron antineutrinos relative to electron neutrinos at the initial point is $\alpha$. For our calculations we adopt $\alpha = 0.8$, $\beta = 0.48$, and $\bar{\beta} =\beta/\alpha = 0.6$ such that x-type neutrinos and antineutrinos are assumed to have equal initial fluxes. These choices are motivated by recent large scale supernova simulations \cite{Summa:2015nyk}. 

The evolution of the neutrino and antineutrino density matrices are governed by the Louiville-von Neumann equations:\footnote{Notice that we consider spatial evolution of a stationary system. For a derivation and listing of underlying assumptions of the utilized approach see Ref.~\cite{Volpe:2013jgr}.}:
\begin{equation}
	{\rm i} \frac{{\rm d} {\rho}}{{\rm d} r} = \left[H, {\rho}\ \right] \, \quad\quad
	{\rm i} \frac{{\rm d} \bar{\rho}}{{\rm d} r} = \left[\bar{H}, \bar{\rho}\ \right] , \\ 
   \label{eq:evoeqs}	
\end{equation}
where $H$ and $\bar{H}$ are the total neutrino and antineutrino Hamiltonians.
At a given location $r$, the survival probabilities, $P_{ij} = P(\nu_i \to \nu_j)$, $\bar{P}_{ij} = P(\bar{\nu}_i \to \bar{\nu}_j)$, can be found given the elements of the density matrices at $r$ and the initial conditions. In particular $\rho_{ee}(r)$ and $\bar{\rho}_{ee}(r)$ are given by 
\begin{equation}
\begin{aligned}
\rho_{ee} &= P_{ee}\,\rho_{ee}(0) + P_{xe}\,\rho_{xx}(0)\\
\bar{\rho}_{ee} &= \bar{P}_{ee}\,\bar{\rho}_{ee}(0) + \bar{P}_{xe}\,\bar{\rho}_{xx}(0)
\end{aligned}
\end{equation}
Using the initial conditions given in Eq.~\eqref{eq:InitialCondition} and for two flavors $P_{xe} = 1-P_{ee}$ and $\bar{P}_{xe} = 1-\bar{P}_{ee}$ we derive the survival probabilities for electron neutrinos and electron antineutrinos to be
\begin{equation}
\label{eq:SurvivalProbabilities}
\begin{aligned}
	(1-\beta)P_{ee} &= (1+\beta)\rho_{ee} - \beta\\
    (1-\bar{\beta})\bar{P}_{ee} &= (1+\bar{\beta})\bar{\rho}_{ee} - \bar{\beta}
\end{aligned}
\end{equation}
Note that we only use the above equation in the cases of $\beta, \bar{\beta} \neq 1$. Because of the way the density matrices have been defined, if $\beta = 1$, this equation requires that $\rho_{ee} = \frac{1}{2}$ everywhere as expected and similarly for the antineutrinos.

The flavor basis neutrino Hamiltonian can be written as
\begin{equation}
\label{eq:HF}
H = \left(
		\begin{array}{cc}
		H_{ee}	& H_{ex} \\ 
		H_{xe} & H_{xx} \\
		\end{array}
	\right) = H_V + V_{\nu}+V_M  \ ,
\end{equation}
where $H_V$ is the vacuum Hamiltonian, $V_{\nu}$ the neutrino-neutrino interaction potential, and $V_M$ the matter potential. The anti-neutrino Hamiltonian is $\bar{H} = H_V -V_{\nu}^{\star}-V_{M}^{\star}$. We shall discuss the terms in this equation shorty but quite generally, if the potentials $V_{\nu}$ and $V_M$ in the Hamiltonian vary with distance $r$ then it is possible for the difference between the diagonal elements of the Hamiltonian to vanish, i.e.\ $H_{ee}-H_{xx}=0$ for neutrinos or $\bar{H}_{ee}-\bar{H}_{xx}=0$ for antineutrinos, leading to a resonance phenomenon. In the case of neutrinos this resonance phenomenon is named after Mikheyev, Smirnov and Wolfenstein (MSW) \cite{PhysRevD.17.2369,Mikheyev:1985aa,Mikheyev:1986tj} and the locations where $H_{ee}-H_{xx}=0$ or $\bar{H}_{ee}-\bar{H}_{xx}=0$ are known as MSW resonances. For non-monotonic potentials it is possible the resonance occurs at multiple locations in which case each resonance is given a name to distinguish it from the others. In the standard case of neutrino oscillations, the H (High) and L (Low) resonances refer to the mass differences $\delta m_{31}$ and $\delta m_{21}$ respectively. Here we will focus on the H resonance only.

The vacuum Hamiltonian for two neutrino flavors is given by
\begin{equation}
H_V = \frac{\delta m^2}{4E}\left(
		\begin{array}{rr}
		-\cos(2\theta_V) & \sin(2\theta_V) \\ 
		\sin(2\theta_V)  & \cos(2\theta_V) 
		\end{array}
	\right) \ , 
    \label{eq:HamVac}
\end{equation}
with $\delta m^2$ the difference between the square of the neutrino masses, $E$ the energy, and $\theta_V$ the mixing angle in vacuum. The neutrino mass splitting used is $\delta m^2 = \pm 2.4 \times 10^{-3}\;{\rm eV}^2$, with a positive sign for the normal hierarchy and a negative sign for the inverted mass hierarchy. The mixing angle we use is $\theta_V = 9^{\circ}\;(0.1571\;{\rm rad})$ and we adopt a single neutrino energy of 20 MeV. In our two flavor calculations we use a mixing angle $\theta_V$ that corresponds to the measured value of $\theta_{13}$ and a mass splitting $\delta m^2$ that that corresponds to $\delta m_{31}^2$. We make this choice so that our two flavor results will include the H resonance which occurs deeper in the star than the L resonance, and so that we can study the effect of the the neutrino mass hierarchy on our results.

The neutrino-neutrino interaction potential for a neutrino emitted from the neutrinosphere at a single angle is given by
\begin{equation}
\label{eq:Vnu}
V_{\nu}(r) = \mu_{\nu} \left( (1+\beta)\rho - \alpha(1+\bar{\beta})\bar{\rho}^{\star} \right) \ ,
\end{equation}
where the star indicates we have taken the complex conjugate of the antineutrino density matrix. The neutrino-neutrino interaction strength is $\mu_{\nu}$ which we take to be 
\begin{equation}
	\mu_{\nu}  = \mu_{0} \left(\frac{r_{\nu}}{r}\right)^4 \ ,
    \label{eq:munu}
\end{equation}
with $r_{\nu} = 10\; {\rm km}$ as the radius of the neutrinosphere, and $\mu_0 = 10^6\;{\rm km^{-1}}$, representing a typical value for the initial relative strength of the interaction \cite{Mirizzi:2015eza}.

The matter potential contains the usual standard contribution plus the NSI: $V_M = V_{MSW} + V_{NSI}$. The standard potential \cite{PhysRevD.17.2369} is 
\begin{equation}
\label{eq:VMSW}
V_{MSW} = \sqrt{2}\,G_{F}\,n_e \left(
		\begin{array}{cc}
		1 & 0 \\ 
		0 & 0 
		\end{array}
	\right) \ ,
\end{equation}
with $G_F$ the Fermi constant, and $n_e$ the net electron number density arising from the difference between the electron and positron number densities: $n_e \equiv n_{e^{-}} - n_{e^{+}}$. The net electron density $n_e$ is also equal to $n_e = Y_e n_N$ where $Y_e$ is the electron fraction and $n_N = n_{\rm p} + n_{\rm n}$ the nucleon density i.e.\ the sum of the densities of protons and neutrons. 
Throughout this paper we adopt a MSW potential of the form $V_{MSW}(r) = \lambda(r)\, Y_e(r)$, where $\lambda(r)$ characterizes the density profile,
\begin{equation}
	\lambda(r) = \sqrt{2} G_F\,n_N(r) = \lambda_0 \left(\frac{r_{\nu}}{r}\right)^3 \ ,
    \label{eq:lmbdae}
\end{equation}
with $\lambda_0 = 10^6\;{\rm km^{-1}}$ as the initial strength of the matter-interaction potential representative of typical densities found in supernovae at $r = r_{\nu}$~\cite{Mirizzi:2015eza,Summa:2015nyk}. For the electron fraction, $Y_e(r)$, we use the same parametrization as described in Esteban-Pretel, Tom\`as and Valle ~\cite{PhysRevD.76.053001}: 
\begin{equation}
\label{eq:Ye}
	Y_e(r) = a + b \tan^{-1}\left(\frac{r - r_{\nu}}{r_s}\right) \ ,
\end{equation} 
and we have set a = 0.308, b = 0.121, $r_s = 42\;{\rm km}$ based upon a fit to the electron fraction at bounce in the $10.8\;M_{\odot}$ simulation by Fischer \emph{et al.} ~\cite{2010A&A...517A..80F}. For $r$ close to $r_{\nu}$ the electron fraction is $Y_e(r)=a=0.308$ while for $r \gg r_s$ the electron fraction has climbed to $Y_e(r) \approx 0.5$

\subsection*{The NSI potential}

The nonstandard interactions are taken to be of a general form of a sum over all fermions present in the matter (ignoring the heavy quark content of the nucleons) and scaled relative to the MSW potential. Thus we write
\begin{equation}
V_{NSI} = \sqrt{2}\,G_F \sum_f n_f\, \epsilon^f  \ ,
\end{equation}
with $f\in \{ e,\,d,\,u \}$ for electrons, down quarks and up quarks respectively. The $\epsilon$'s are Hermitian matrices with elements describing the strengths of the nonstandard interactions. The NSI potential can be rewritten by introducing the fermion fraction $Y_f$ defined to be 
\begin{equation}
Y_f \equiv \frac{n_f}{n_{\rm N} } \ ,
\end{equation}
Assuming charge neutrality of the medium, the fermion fractions for the down quark and up quark can be expressed in terms of the electron fraction, $Y_e$, as:
\begin{equation}
	\begin{aligned}
		Y_d &= 2 - Y_e \ , \\
		Y_u &= 1 + Y_e \ .
	\end{aligned}
\end{equation}
The NSI potential is thus 
\begin{eqnarray}
V_{NSI} & = & \sqrt{2}\,G_F\,n_N \left( Y_e\,\epsilon^e + (1+Y_e)\,\epsilon^u + (2-Y_e)\,\epsilon^d\right)
\nonumber \\ & & \\
 & = & \lambda(r) \left( Y_e\,\epsilon^e + (1+Y_e)\,\epsilon^u + (2-Y_e)\,\epsilon^d\right). \label{eq:VNSI}
\end{eqnarray}
It was shown by Maltoni \& Smirnov \cite{2016EPJA...52...87M} and more recently by Coloma \& Schwetz \cite{2016arXiv160405772C} that oscillation data alone provide very poor if any constraint on the NSI parameters. Strong constraints only emerge when scattering experiments are included. From a combination of terrestrial and solar neutrino oscillation and scattering experiments, upper limits have been placed upon the NSI parameters \cite{2003JHEP...03..011D,2007JHEP...12..002C,2013RPPh...76d4201O,2009JHEP...08..090B}. The model independent constraints from Biggio, Blennow and Fernandez-Martinez~\cite{2009JHEP...08..090B} make no assumption about the origin of the NSI. The constraints are not upon the individual coupling of the neutrinos to each particular fermion but rather they define an effective NSI coupling to matter, $\epsilon^{mat}$, as
\begin{equation}
\label{eq:epsm}
\epsilon^{mat} = \sum_f \frac{n_f}{n_e} \epsilon^f = \sum_f \frac{Y_f}{Y_e} \epsilon^f \ .
\end{equation}
For Earth like matter, assuming equal numbers of neutrons and protons and electrons, the constraints for the elements of $\epsilon^{mat}$ given by Biggio, Blennow, and Fernandez-Martinez are
\begin{equation}
\label{eq:NSIConEarth}
\left(\begin{array}{lll}
		|\epsilon_{ee}| < 4.2 & |\epsilon_{e\mu}| < 0.33 & |\epsilon_{e\tau}| < 3.0\\ 
		 & |\epsilon_{\mu\mu}| < 0.068 & |\epsilon_{\mu\tau}| < 0.33 \\ 
         		 &  & |\epsilon_{\tau\tau}| < 21 
		\end{array}
	\right) \ .
\end{equation}
For `solar like' matter, consisting only of protons and electrons, their constraints are 
\begin{equation}
\label{eq:NSIConStar}
\left(\begin{array}{lll}
		|\epsilon_{ee}| < 2.5 & |\epsilon_{e\mu}| < 0.21 & |\epsilon_{e\tau}| < 1.7 \\ 
		 & |\epsilon_{\mu\mu}| < 0.046 & |\epsilon_{\mu\tau}| < 0.21 \\ 
         		 &  & |\epsilon_{\tau\tau}| < 9.0
		\end{array}
	\right) \ .
\end{equation}
We see that, except for $\epsilon_{\mu\mu}$, the current experimental constraints on NSI parameters are remarkably loose. 

Even though large NSI effects for solar neutrinos are possible if one is prepared to adjust the mass splitting and mixing angles, we shall adopt a conservative approach and maintain the standard Mikheyev \& Smirnov \cite{Mikheyev:1985aa,Mikheyev:1986tj} solution for the solar neutrino problem by requiring the NSI in the Sun to be small. This choice does not mean the NSI in supernovae are also small: as Eq.~(\ref{eq:VNSI}) shows, the effect of the neutrino NSI depends upon the composition of the matter. In supernovae or compact object mergers the electron fraction can become much smaller than in the Sun permitting the NSI to be significant. The requirement that the NSI vanish when the electron fraction $Y_e$ is the solar electron fraction $Y_{\odot}$ leads from Eq.~(\ref{eq:VNSI}) to the following condition on the $\epsilon$ parameters:
\begin{equation}
0 = Y_{\odot}\, \delta \epsilon^{e} + (1+Y_{\odot})\,\delta \epsilon^{u}+ (2-Y_{\odot})\,\delta \epsilon^{d} \ ,
\label{eq:NSIzero}
\end{equation}
where $\delta \epsilon^{f} = \epsilon_{ee}^{f} -\epsilon_{xx}^{f}$. The value for $Y_{\odot}$ is the electron fraction at the MSW resonance in the standard oscillation case ($Y_\odot \approx 0.7$). Eq.~(\ref{eq:NSIzero}) implies a relationship between the difference between one set of NSI parameters, e.g.\ $\delta \epsilon^{e}$, in terms of the other differences $\delta \epsilon^{u}$ and $\delta \epsilon^{d}$ for any given choice of solar electron fraction. We solve Eq.~(\ref{eq:NSIzero}) for $\delta \epsilon^{e}$ and substitute into Eq.~(\ref{eq:VNSI}).  For ease of calculation, we also set the off-diagonal elements of $\epsilon^e$, $\epsilon^u$ and $\epsilon^d$ to  $\epsilon^{e}_{ex} = \epsilon^{u}_{ex} = \epsilon^{d}_{ex} \equiv \epsilon_0$. Putting everything together we write our NSI potential as  
\begin{equation}
V_{NSI} = \lambda(r) \,\left(
		\begin{array}{cc}
		\left(\frac{Y_{\odot} - Y_e}{Y_{\odot}} \right)\,\delta\epsilon^{n} & (3+Y_e)\,\epsilon_0 \\ 
		(3+Y_e)\,\epsilon_0^{*} & 0 
		\end{array}
	\right) \ .
    \label{eq:newVNSI}
\end{equation}
The factor of $(3+Y_e)$ comes from the inclusion of nonstandard couplings to the up and down quarks as well as the electrons. If only the coupling to electrons had been included this factor would be $Y_e$. A term proportional to a unit matrix has been subtracted in order to zero the lower diagonal element and we have rewritten the combination $\delta \epsilon^{u} + 2\,\delta\epsilon^{d}$ as the NSI coupling to the neutron $\delta \epsilon^{n} = \delta \epsilon^{u} + 2\,\delta\epsilon^{d}$. From hereon we shall use $\delta \epsilon^n$ and $\epsilon_0$ as the NSI parameters. 

The limits in Eqs. (\ref{eq:NSIConEarth}) and (\ref{eq:NSIConStar}) can be directly translated to limits on $\delta\epsilon^n$ and $\epsilon_0$. For a given electron fraction $Y_e$, from Eqs. (\ref{eq:VNSI}) and (\ref{eq:epsm}) we find 
\begin{equation}
|\epsilon_0| < \left(\frac{Y_e}{3+Y_e}\right)|\epsilon_{ij}^{mat}|.
\end{equation}
The $\epsilon_{ij}$ is chosen as appropriate to the type of two flavor calculations. For the purposes of this paper we use $\epsilon_{e\tau}$. Similarly for $\delta\epsilon^n$ we find
\begin{equation}
|\delta\epsilon^n|<\left(\frac{Y_e Y_{\odot}}{Y_{\odot}-Y_e}\right) |\delta\epsilon^{mat}|
\end{equation}
where $\delta\epsilon^{mat}$ is the difference between the diagonal elements of the effective matter coupling defined in equation (\ref{eq:epsm}). Since we compute $e-\tau$ mixing, the limits on our NSI parameters stem from a $\delta\epsilon^{mat} = \epsilon^{mat}_{ee}-\epsilon^{mat}_{\tau\tau}$. Thus we find the limits for $\epsilon_0$ are of order ${\cal{O}}(0.1-1)$ and $\delta\epsilon^n$ are of order ${\cal{O}}(1-10)$.
\begin{figure}
\includegraphics[width=0.47\textwidth]{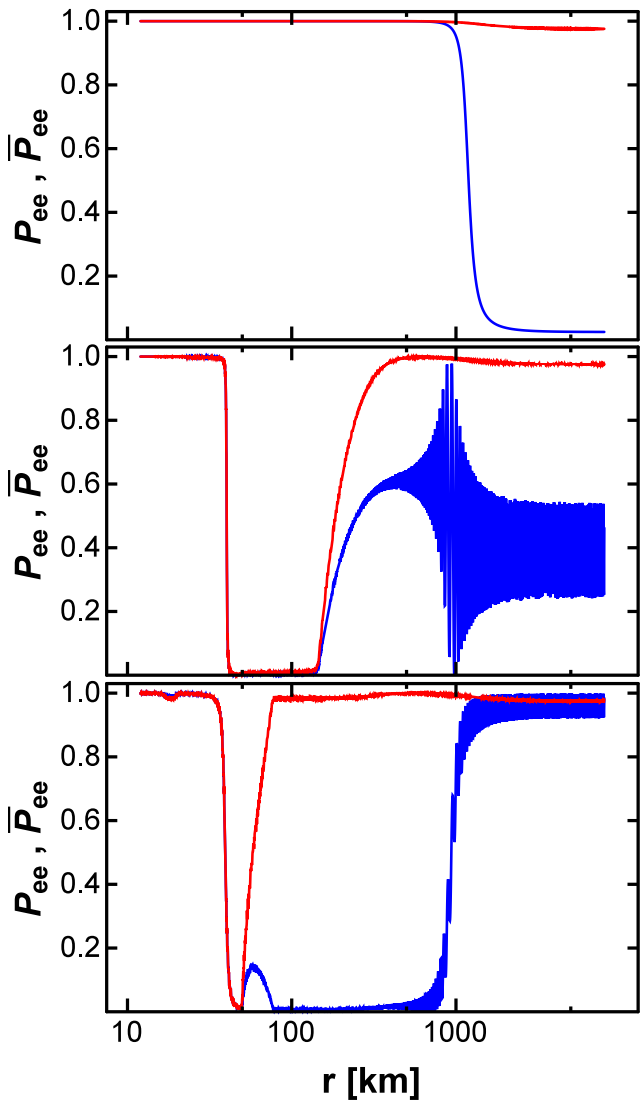}
\caption{Normal hierarchy survival probabilities, {\rm $P_{ee} = P(\nu_e \to \nu_e)$, $\bar{P}_{ee} = P(\bar{\nu}_e \to \bar{\nu}_e)$} for electron neutrinos (blue) and antineutrinos (red). The top figure shows the results of our calculations in the absence of any NSI. The middle figure includes NSI terms with the parameters set to $\delta\epsilon^n = -0.84$ and $\epsilon_0 = 0.00025 $ and one observes an I resonance at $r\sim 40\;{\rm km}$ and a bipolar/nutation oscillations beginning at $r\sim 150\;{\rm km}$. Finally, in the bottom figure we increase $\epsilon_0$ to 0.001 and observe that the I resonance is now followed by a new type of transition which we shall show is a Standard MNR.}
\label{fig:BigResult}
\end{figure}

\subsection*{NSI effects}

To prepare for future discussions we illustrate some of the effects of the NSI in figure (\ref{fig:BigResult}). The figure shows the results from three calculations for the electron neutrino and antineutrino survival probabilities (Eq.~\ref{eq:SurvivalProbabilities}) as a function of distance. The neutrino mass hierarchy in each case is chosen to be normal. In the top panel is the case of no NSI and one observes no flavor transformation until the neutrino reaches $r\sim 1000\;{\rm km}$ which is the location of the MSW H resonance. In the middle panel we switch on the NSI using $\delta\epsilon^n = -0.84$ and $\epsilon_0 = 2.5\times10^{-4}$ and find a result similar to those of Esteban-Pretel, Tom{\`a}s, and Valle \cite{2010PhRvD..81f3003E}. One observes a NSI-induced flavor conversion at $r\sim 40\;{\rm km}$ (which  is the I resonance) and then another effect not usually  seen in seen in the normal hierarchy which starts at $r\sim 150\;{\rm km}$ which we will show is the bipolar/nutation oscillations. In the bottom panel we have used a different value for the NSI parameter $\epsilon_0$ set to $\epsilon_0 = 0.001$ and find something completely different than the two panels above. The transformation at $r\sim 40\;{\rm km}$ is followed by a new transformation - which we shall show is a matter-neutrino resonance \cite{2012PhRvD..86h5015M,2014arXiv1403.5797M,2015arXiv151000751V,2016PhLB..752...89W}. 

In addition to the two examples of the effects of NSI for supernovae neutrinos shown in the two lower panels we have undertaken many thousands of similar calculations exploring the NSI parameter space. Before we present further representative cases and partition the NSI parameter space according to which effects are observed, we describe analytically the various transformation effects we have found. 

\section{Analytical Description}
\label{sec:Matter}

\subsection{I resonances}

\begin{figure}[b]
\includegraphics[clip,width=0.47\textwidth]{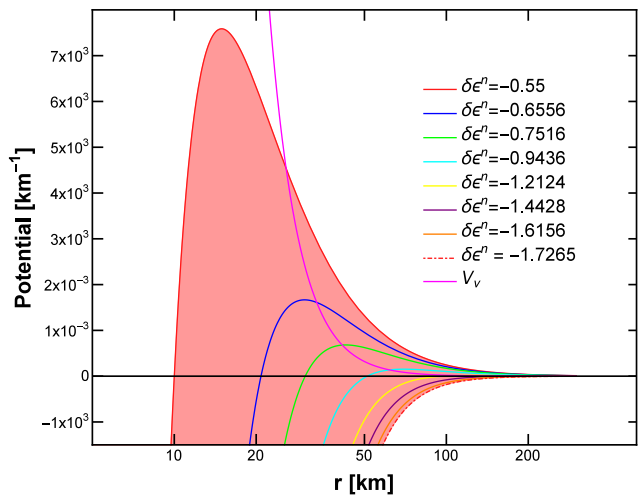}
\caption{The non-zero diagonal element of the total matter potential $V_M = V_{MSW} + V_{NSI}$ as a function of $r$ for eight different values of $\delta\epsilon^n$, and the $V_{\nu}$ scaling parameter, $\mu_{\nu}$. The shaded red region is set by the requirement the I resonance occurs entirely outside the neutrinosphere.}
\label{fig:VNSIvR}
\end{figure}

One of the consequences of the NSI can be new MSW resonances. In particular, the first obvious feature seen in the lower panels of figure (\ref{fig:BigResult}) is an I resonance at $r\sim 40\;{\rm km}$. These new resonances emerge due to the modification of the matter contribution to the Hamiltonian. Together the MSW and the NSI potentials form the total matter potential $V_M$ and the way we have written both potentials means $V_M$ has only one non-zero element on the diagonal.
\begin{equation}
\begin{aligned}
V_M &= V_{MSW} + V_{NSI}\\ &= \lambda(r)\begin{pmatrix} Y_e(r) + \delta\epsilon^n\left(\frac{Y_{\odot} - Y_e(r)}{Y_{\odot}}\right) & (3+Y_e(r))\epsilon_o \\ (3+Y_e(r))\epsilon_o & 0 \end{pmatrix}
\end{aligned}
\end{equation}
This element is a function of the electron fraction $Y_e$ and with certain combinations of the NSI parameters one finds it is possible for the total matter potential to have a different sign at different values of $Y_e$. Several examples of this evolution of the total matter potential can be seen in figure (\ref{fig:VNSIvR}) where we plot the diagonal term of the total matter potential as a function of radius $r$ for various values of $\delta\epsilon^n$. As a result of our imposed constraint on NSI effects in the Sun, the diagonal term of $V_M$ must be positive at large $r$/high $Y_e$. As one moves toward smaller $r$, the densities increase causing $\lambda(r)$ to increase, while the $Y_e$ decreases. This causes the diagonal component of the total matter potential to peak at some maximum positive value and then fall through zero at $r = r_0$ and become negative.  

The location where the diagonal element of the matter potential changes sign is determined by finding the location $r$ which gives an electron fraction that satisfies
\begin{equation}
Y_e + \delta\epsilon^n \left(\frac{Y_{\odot} - Y_e}{Y_{\odot}} \right) = 0 \ .
\label{eq:zerocrossing}
\end{equation}
If we solve this equation for $Y_e$ then we find the total matter potential will be negative when the electron fraction is less than 
\begin{equation}
\label{eq:Y0}
Y_e < - \frac{\delta\epsilon^{n}\,Y_{\odot} }{Y_{\odot} -\delta\epsilon^{n} } \equiv Y_{0} \ .
\end{equation}
If $\delta\epsilon^{n} < 0$ then wherever the electron fraction $Y_e$ is below the threshold $Y_{0}$, the matter potential is negative. We stress that this occurs without greatly affecting solar neutrinos. If $Y_\odot \approx 0.7$ and we consider a range of $\delta \epsilon^n \in [-0.5,-2.0]$ then we find the range of $Y_0$ that allows for this cancellation is $Y_0 \in [0.292,0.519]$, which overlaps significantly with the electron fractions typically found in supernovae simulations, as fit in Eq.~\eqref{eq:Ye}. Note that as $\delta\epsilon^{n}$ becomes increasingly negative, $Y_0$ becomes increasingly positive causing the location of $r_0$ to move outwards. 

The position of the I resonance, $r_I$, has been defined by Esteban-Pretel, Tom{\`a}s, and Valle \cite{PhysRevD.76.053001} by setting the two diagonal elements of the neutrino Hamiltonian equal i.e. $H_{ee} = H_{xx}$ neglecting the neutrino-neutrino interaction. Thus for neutrinos the position of the I resonance, $r_I$, is defined to be the location where
\begin{equation}
\label{eq:rI}
\frac{\delta m^2}{2E} \cos2\theta_V = \lambda(r_I)\left[ Y_e(r_I) + \delta\epsilon^n \left(\frac{Y_{\odot} - Y_e(r_I)}{Y_{\odot}} \right)\right]
\end{equation}
with a relative negative sign needed to predict $\bar{r}_I$, the location of the I resonance for antineutrinos. We have verified that within the NSI parameter space considered here, the neutrino-neutrino interaction has a negligible effect on the position, width, and adiabaticity of the transformation that occurs at the I resonance. 

As seen in figure~(\ref{fig:VNSIvR}), the location where the non-zero diagonal component of $V_M = 0$ is very close to the location where this component is equal to the vacuum scale in either the normal or inverted hierarchies. We can therefore approximate the location of the I resonance by setting the vacuum term on the left had side of Eq.~\eqref{eq:rI} equal to 0. Using this approximation we predict the location of the I resonance is the same for both neutrinos and antineutrinos in both the normal and inverted hierarchy, $r_I \approx \bar{r}_I \approx r_0$. This approximation holds best for small values of $|\delta\epsilon^n|$ when the location of the I resonance is close to the proto-neutron star. Figure~(\ref{fig:VNSIvR}) also shows how the slope of the diagonal component of $V_M$ at the location of $r_0$ decreases as $|\delta\epsilon^n|$ increases. This will cause the distance between $r_I$, $\bar{r}_I$ and $r_0$ to increase. 
\begin{figure}[b]
\centering
\includegraphics[width=0.48\textwidth]{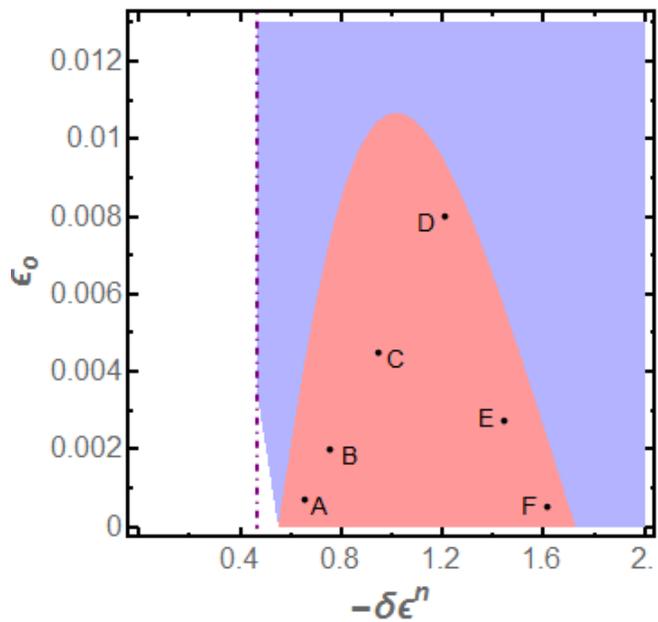}
\caption{Outline of the parameter space considered in this paper. The magenta region represents areas where we expect to see NSI effects due to the I resonance. The white region represents areas where our model is not applicable. The blue region contains uncertain NSI effects where the neutrinosphere and I resonance overlap. The dashed purple line shows the location where the solution for the I resonance becomes nonphysical i.e.\ $r_I \leq 0$. The six dots in the red region represent the six test cases shown in figures~(\ref{fig:Bipolar1Case}),~(\ref{fig:MNRCase}),~(\ref{fig:?MNRCase}),~(\ref{fig:Bipolar2Case}),~(\ref{fig:ChaoticCase}), and~(\ref{fig:NoMSWCase}) from left to right.}
\label{fig:ParamOverview}
\end{figure}

The I resonance has a width. We can define this width by first finding the eigenvalues $\tilde{k}_i$ of the total Hamiltonian, given by 
\begin{equation}
\tilde{k}_i = \frac{H_{ee}+H_{xx}}{2} \pm \frac{1}{2}\sqrt{\left(H_{ee}-H_{xx}\right)^2 + 4\,|H_{ex}|^2 } \ ,
\end{equation}
and the matter mixing angle $\tilde{\theta}$ is defined to be 
\begin{equation}
\tan^2 \tilde{\theta} = \frac{H_{ee}-\tilde{k}_1}{H_{xx}-\tilde{k}_1} \ ,
\label{eq:Qtilde}
\end{equation}
If we Taylor expand the function $\sin^2\,(2\tilde{\theta})$ around the resonance we find 
\begin{equation}
\sin^2(2\tilde{\theta}) \approx 1 - 4\,\left(\left.\frac{d\tilde{\theta}}{dr}\right|_{r_I}\right)^2\,(\delta r)^2 + \ldots \ ,
\end{equation}
then we define the width $\sigma_I$ to be 
\begin{eqnarray}
\sigma_I & = & \left( \left.\frac{d\tilde{\theta}}{dr}\right|_{r_I} \right)^{-1} \\
& = & \left(\left[\frac{1}{4\sqrt{|H_{ex}|^2}} \left(\frac{dH_{ee}}{dr} - \frac{dH_{xx}}{dr}\right)\right]_{r_I}\right)^{-1} \ .
\end{eqnarray}
where $\left.\frac{d\tilde{\theta}}{dr}\right|_{r_I}$ is defined using Eq.~\eqref{eq:Qtilde}.

The position and width of the I resonance as a function of the NSI parameters partitions the NSI parameter space into three regions: 
\begin{itemize}
\item $r_I + \sigma_I \leq r_{\nu}$: the I resonance is entirely inside the neutrinosphere represented by the white region in figure (\ref{fig:ParamOverview})
\item $r_I - \sigma_I \leq r_{\nu} \leq r_I + \sigma_I$: the neutrinosphere and I resonance overlap, represented by the blue region in figure (\ref{fig:ParamOverview})
\item $r_{\nu} \leq r_I - \sigma_I$: the I resonance is beyond the neutrinosphere, represented by the red region in figure (\ref{fig:ParamOverview}) 
\end{itemize}
Given the setup of our calculations, in particular that the neutrinos are free streaming, we can only reliably calculate the effects of NSI when the I resonance is beyond the neutrinosphere. For this reason we will focus our attention in the rest of our paper upon this third region. The effects of NSI with parameters outside the third region would require the use of different methods for neutrino transport. 

Finally, knowing the width of the I resonance allows us to determine the adiabaticity of the resonance since the adiabaticity is determined by the ratio of the width compared to the oscillation length at the resonance $\ell_I$. The oscillation length is 
\begin{equation}
\ell_I = \frac{2\pi}{\sqrt{|H_{ex}(r_I)|^2}} \ ,
\end{equation}
so the adiabaticity, $\gamma_I$, given by the ratio $\gamma_I = \sigma_I / \ell_I$, is 
\begin{equation}
\gamma_I = \left[\frac{2|H_{ex}|^2}{\pi}\left(\frac{dH_{ee}}{dr} - \frac{dH_{xx}}{dr}\right)^{-1}\right]_{r_I} \ .
\label{eq:adiabaticity}
\end{equation}
If $\gamma_I$ is much greater than unity the evolution is adiabatic and the neutrinos follow the instantaneous - matter - eigenstates. If $\gamma_I$ is less than unity then the evolution is non-adiabatic and the neutrinos jump from following one eigenstate before the resonance to following the other after. The adiabticity of the I resonance depends on the gradients of the potentials, as well as the size of the off-diagonal element of the Hamiltonian. The off-diagonal elements enter in the numerator so that the I resonance becomes more adiabatic as $|H_{ex}|$ increases. Using Eq.~\eqref{eq:HamVac} and Eq.~\eqref{eq:newVNSI} we can see that the NSI will dominate this term for $\epsilon_0 \gtrsim 10^{-5}$. 

\subsection{H resonance}

The high density resonance \cite{2000PhRvD..62c3007D} is seen with standard neutrino oscillation physics alone and it appears in all three panels of figure (\ref{fig:BigResult}) occurring around $r\sim 1000\;{\rm km}$.  Like the I resonance, the H resonance is a MSW transition and therefore it occurs at a location where Eq.~\eqref{eq:rI} is also true. The H resonance is typically not near the zero crossing $r_0$: it arises because the function $\lambda(r)$ - given in Eq. (\ref{eq:lmbdae}) - decreases with distance. The NSI do still affect the H resonance nevertheless. The location of the H resonance, $r_H$, is pushed inwards and toward higher density compared to the position for no NSI when $\delta\epsilon^n <0$, and outwards and to lower density compared to the position for no NSI when $\delta\epsilon^n >0$. However this relative change in position of the H resonance does not lead to any change in the flavor survival probabilities at the edge of the supernova. The adiabaticity, $\gamma_H$, of the H resonance, defined by evaluating Eq.~\eqref{eq:adiabaticity} using the appropriate matrix elements and their derivatives at $r_H$, remains high even with the NSI contribution. This means that the neutrinos and antineutrinos follow the eigenstates and undergo a significant change in flavor as a result of the interactions with the high matter density. 

\subsection{MNR}

As seen in figure (\ref{fig:BigResult}), one new effect that emerges for supernova neutrinos with NSI is the Matter Neutrino Resonance. 
A Matter Neutrino Resonance occurs when the background matter contribution to the neutrino Hamiltonian cancels with the contribution from the neutrino-neutrino interaction. In merger scenarios, with SM physics, a cancellation can occur close to the neutrino emission region even if the matter potential always remains positive because the antineutrino flux can dominate over the neutrino flux resulting in a negative neutrino-neutrino potential \cite{2012PhRvD..86h5015M,2014arXiv1403.5797M,2015arXiv151000751V,2016PhLB..752...89W}. In supernovae (with spherically symmetric emission) neutrino fluxes always dominate over antineutrinos and, as such, the neutrino-neutrino potential always remains positive using only SM Physics. MNR conditions are, therefore, not seen in supernovae. 

With NSI, however, the matter potential can become negative, as shown in figure~(\ref{fig:VNSIvR}), or, as a consequence of the I resonance, the diagonal component of the neutrino-neutrino interaction can become negative. The former leads to Symmetric MNR, the latter leads to a Standard MNR \cite{2014arXiv1403.5797M,2016PhRvD..93d5021M}. 

For either the Standard or Symmetric MNR the system starts with a relative sign between the diagonal part of the neutrino-neutrino interaction and the diagonal contribution from the total matter potential, and the potentials are arranged such that $|V_{\nu}| > |V_M|$. From Eq.~\eqref{eq:munu}, the neutrino-neutrino potential falls off as $1/r^4$ while $V_M\propto \delta\epsilon^n/r^3$ - Eqs.~\eqref{eq:newVNSI} and \eqref{eq:lmbdae}. Thus for certain values of $\delta\epsilon^n$ we can generate a cancellation where $|V_{\nu}| = |V_M|$. The conversion of neutrinos and antineutrinos can cause $V_{\nu}$ to change in such as way as to maintain that cancellation over a finite distance. In a Symmetric MNR, both neutrinos and antineutrinos transform in a similar manner while in the Standard MNR neutrinos and antineutrinos transform asymmetrically ending up with different final flavor configurations. The bottom panel of figure~(\ref{fig:BigResult}) illustrates an example case in which we can see both types of MNR transitions induced by NSI. The Symmetric MNR is evident as the first small dip at a few tens of kilometers and the Standard MNR occurs further out at around 50 km.

The required cancellations outlined above come from the MNR condition in which $H_{ee}\approx H_{xx}$ where we have neglected the vacuum contribution~\citep{2015arXiv151000751V}:
\begin{eqnarray}
0&\approx& \lambda Y_e + \lambda\delta\epsilon^n\left(\frac{Y_{\odot}-Y_e}{Y_{\odot}}\right) \nonumber \\ 
& & + \mu_{\nu}\left((1+\beta)(\rho_{ee}-\rho_{xx}) - \alpha (1+\bar{\beta})(\bar{\rho}_{ee} - \bar{\rho}_{xx})\right) \nonumber \\ 
&&
\end{eqnarray}
In Malkus, Friedland and McLaughlin \cite{2014arXiv1403.5797M} it was shown how this understanding leads to analytic expressions for electron (anti)neutrino survival probabilities during MNR transitions. With an initial flux of only electron and anti-electron type neutrinos
\begin{equation}
\label{eq:OldMNR}
\begin{aligned}
	P_{ee} &=  \frac{1}{2} \left(1 + \frac{\alpha^2-1 - R^2}{2 R} \right), \\
	\bar{P}_{ee} &= \frac{1}{2} \left(1 + \frac{\alpha^2-1 + R^2}{2 \alpha R} \right) ,
\end{aligned}
\end{equation}
where $R \equiv V_{\rm MSW}/\mu_{\nu}$ is the ratio of the neutrino-electron and neutrino-neutrino interaction scales, and $\alpha$ is the initial asymmetry between electron neutrinos and antineutrinos. We have a second flavor of neutrino and antineutrino present initially and we also take into account the NSI. Following the same steps - detailed in the appendix - we find the more general expression
\begin{equation}
\label{eq:MNRPred}
	\begin{aligned}
	P_{ee} &=  \frac{1}{2} \left(1 + \frac{\alpha^2(1-\bar{\beta})^2-(1-\beta)^2 - q^2}{2(1-\beta) q} \right)  \ , \\
	\bar{P}_{ee} &= \frac{1}{2} \left(1 + \frac{\alpha^2(1-\bar{\beta})^2-(1-\beta)^2 + q^2}{2\alpha(1-\bar{\beta}) q} \right) ,
\end{aligned}
\end{equation}
where we have replaced R with q to include the NSI effect to the total matter potential: $q \equiv (V_{\rm MSW} + V_{\rm NSI})/\mu_{\nu}$. Notice that the expressions in Eq.~\eqref{eq:MNRPred} obey the same conditions as the survival probabilities in Eq.~\eqref{eq:SurvivalProbabilities}. The new terms in Eq.~\eqref{eq:MNRPred} compared to Eq.~\eqref{eq:OldMNR} account for the flux of x type (anti)neutrinos via the inclusion of the flavor asymmetries $\beta$ and $\bar{\beta}$. Note that the flavor evolution through both types of MNR are adiabatic (see Ref.~\cite{2015arXiv151000751V}). We will make use of these expressions in our analysis of the numerical results in order to diagnose the transitions we observe.

\subsection{Bipolar/Nutation}

Finally we discuss the bipolar/nutation type of flavor transition seen in the middle panel of figure (\ref{fig:BigResult}) starting at around 150 km in which antineutrinos (red) fully convert and neutrinos (blue) partially convert. 
Bipolar/nutation type flavor transitions are seen with standard neutrino oscillation physics alone and, like the H resonance, they too can be affected by NSI. 

The region where the neutrinos undergo nutation type transitions can be predicted via linear stability analysis. During a bipolar/nutation transition the off-diagonal elements of the neutrino and antineutrino density matrices grow exponentially i.e.\ they are `unstable' \cite{Hannestad:2006nj}. By applying a linearization procedure \cite{2011PhRvD..84e3013B,2015arXiv151000751V} we can derive the following stability matrix ${\bf \Sigma}$ applicable for pure flavor states in our NSI supernova model:
\begin{equation}
\label{eq:stab22}
{\bf \Sigma} = \left(
	\begin{array}{cc}
		-\displaystyle{\frac{\delta m^2}{2 E}} \mp (1 - \beta) \mu_{\nu} & \pm(1 - \beta) \mu_{\nu} \\
		\mp(\alpha - \beta) \mu_{\nu} & \displaystyle{\frac{\delta m^2}{2 E}} \pm (\alpha - \beta) \mu_{\nu}
	\end{array}
		\right) \ ,
\end{equation}
where upper signs refer to initial conditions set in Eq.~\eqref{eq:InitialCondition} while lower signs are applicable in case of a prior complete flavor conversion. The eigenvalues of this matrix describe the evolution of the off-diagonal elements of the density matrix. When the eigenvalues of the stability matrix become complex the off-diagonal elements of the density matrix grow exponentially. Conversely, the real eigenvalues of the stability matrix correspond to the collective coherent small amplitude oscillation frequencies indicating no substantial flavor evolution. For the stability matrix given in Eq. (\ref{eq:stab22}), complex eigenvalues are obtained when
\begin{equation}
\mu_{\nu}^2\left[1-(1-\beta)(\alpha-\beta)\right] + \mu_{\nu} \frac{\delta m^2}{E} (1-\beta) + \left(\frac{\delta m^2}{2E}\right)^2 < 0.
\label{eq:stabcond}
\end{equation}
Notice that the location of the instability region is independent of the NSI parameters. 

\section{Numerical Results}
\label{sec:effects}

\begin{figure}[t]
\includegraphics[width=0.48\textwidth]{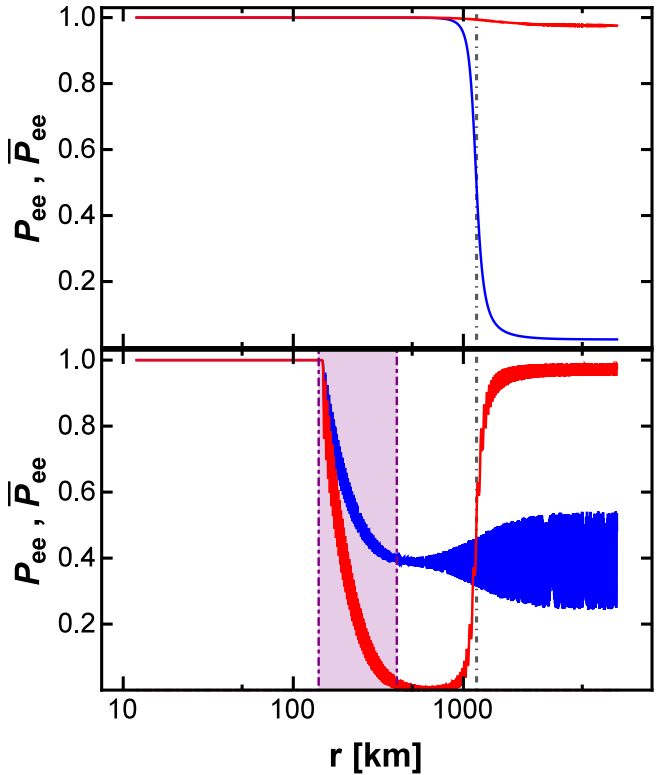}
\caption{Survival probabilities of electron neutrinos (blue) and antineutrinos (red) for the normal (top) and inverted (bottom) hierarchies in the absence of NSI. The vertical shaded band indicates the region where linear stability analysis predicts a bipolar/nutation transformation should occur due to neutrino-neutrino interaction. 
The position of the H resonance for neutrinos in the normal hierarchy and for antineutrinos in the inverted hierarchy is shown as a vertical dash-dotted line at $r_H \approx 1200 \;{\rm km}$.}
\label{fig:NoNSI}
\end{figure}

\begin{figure}[t]
\includegraphics[width=0.47\textwidth]{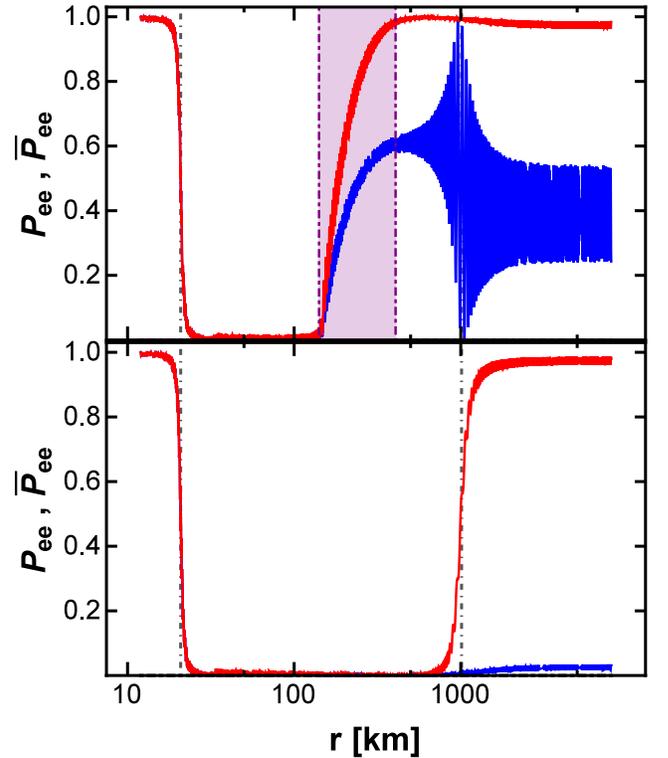}
\caption{The same as in figure (\ref{fig:NoNSI}) but with NSI parameters set to $\delta\epsilon^n = -0.6556$ and $\epsilon_0 = 0.0007$, corresponding to point A in figure~(\ref{fig:ParamOverview}). Again the shaded band indicates where linear stability analysis predicts a bipolar/nutation transformation should occur due to neutrino-neutrino interaction. The vertical gray dot-dashed line at $r\approx 20\;{\rm km}$ is the predicted location of the I resonance according to Eq.~(\ref{eq:rI}) and the vertical dot-dashed line at $r\approx 1000\;{\rm km}$ is the H resonance for neutrinos or antineutrinos.}
\label{fig:Bipolar1Case}
\end{figure}
\begin{figure}[t]
\includegraphics[width=0.47\textwidth]{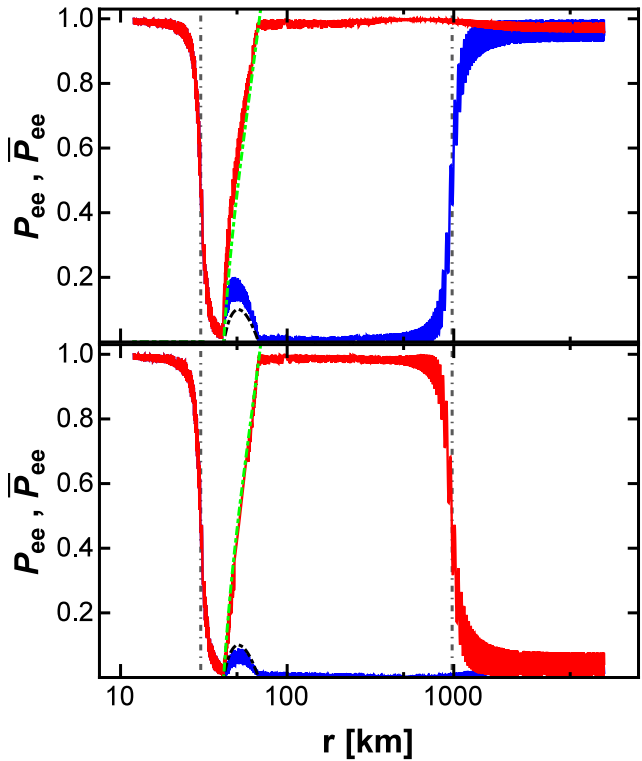}
\caption{The same as in figure (\ref{fig:Bipolar1Case}) but with NSI parameters set to $\delta\epsilon^n = -0.7516$ and $\epsilon_0 = 0.002$ corresponding to point B in figure~(\ref{fig:ParamOverview}). The vertical gray dot-dashed line at $r\approx 30\;{\rm km}$ is the predicted location of the I resonance according to Eq.~(\ref{eq:rI}) and the vertical dot-dashed line at $r\approx 1000\;{\rm km}$ is the H resonance for neutrinos or antineutrinos. At this combination of NSI parameters a MNR occurs starting at $r\sim 40\;{\rm km}$ and the analytical prediction for the evolution of the transition probabilities using Eq.~(\ref{eq:MNRPred}) is shown as the black and green dashed lines for electron neutrinos and antineutrinos respectively.}
\label{fig:MNRCase}
\end{figure}

The analytical tools described in the previous section allow us to identify which type of transition we are observing. Armed with these equations, we shall examine some selected combinations of the NSI parameters but first, we study the case with no NSI effects for reference before we examine six samples of the parameter space shown in figure~(\ref{fig:ParamOverview}). 

The results for the survival probabilities for the electron neutrinos and antineutrinos, $P_{ee}$ and $\bar{P}_{ee}$ in the absence of NSI are shown in figure (\ref{fig:NoNSI}). Note the upper panel of this figure (normal hierarchy) was shown previously in figure~(\ref{fig:BigResult}) and we reproduce it here for convenience. 
The predicted location of the H resonance given by Eq.~\eqref{eq:rI} matches a transition seen in the numerical results and similarly we observe in the inverted hierarchy a transition starting at $r \approx 150\;{\rm km}$ and finishing at $r \approx 400\;{\rm km}$ which matches the shaded region where the linear stability analysis indicates a bipolar/nutation should lie as predicted by Eq.~\eqref{eq:stabcond}. Thus we are confident of our assignments of the two transitions as being bipolar/nutation and an H resonance.  

With figure~(\ref{fig:NoNSI}) as a reference for regular SM physics, let us now switch on the NSI. The six example cases discussed below correspond to the six points in figure (\ref{fig:ParamOverview}), displaying their position in the parameter space. For each point, the total matter potential, $V_M$, is plotted in figure (\ref{fig:VNSIvR}). 

Our first example case, labeled point A in figure~(\ref{fig:ParamOverview}), is for $\delta\epsilon^n = -0.6556$, $\epsilon_0 = 0.0007$. The results for the electron (anti)neutrino survival probabilities as a function of distance for this case are shown in figure (\ref{fig:Bipolar1Case}). The reader will observe not only an H resonance - which is now at $r\approx 1000\;{\rm km}$ due to NSI contributions - but also a number of flavor changing effects that are not present in the previous figure where NSI were absent:
\begin{itemize}
\item an I resonance at $\sim 20$ km, 
\item in the normal hierarchy (top panel), a nutation/bipolar transition starting at $\sim$ 150 km,
\item in the inverted hierarchy (bottom panel), the nutation/bipolar transition has disappeared.
\end{itemize}
The I resonance at $\sim 20$ km leads to a complete swap of the $e$ and $x$ flavors for both neutrinos and antineutrinos indicating the resonance is adiabatic. But as a consequences of the NSI induced I resonance, the stability of the system to collective transformations is also swapped. Now it is the normal hierarchy which is seen to experience a bipolar/nutation transition while for the inverted hierarchy no collective effects occur. Thus effects from NSI can `spill over' and lead to other types of flavor transformations that did not occur in the absence of NSI or switch off transformation that did occur when we only had standard neutrino oscillation physics as seen in figure (\ref{fig:NoNSI}). As before, the location of the H resonance in the numerical results matches the prediction of an observed transition and the new I resonance location also matches an observed transition at that location. Finally, the region where the nutation/bipolar transition occurs for the normal hierarchy again overlaps with the shaded region which is the range in $r$ where the linear stability analysis indicates it should occur. 

Our next example case, point B in figure~(\ref{fig:ParamOverview}), is $\delta\epsilon^n = -0.7516$, $\epsilon_0 = 0.002$ and the survival probabilities for this case are shown in figure (\ref{fig:MNRCase}). When compared to figures (\ref{fig:NoNSI}) and (\ref{fig:Bipolar1Case}), we now observe transformations which did not occur in either of those figures. Immediately after the I resonance located around $r_I \sim 30\;{\rm km}$, the neutrinos undergo a Standard Matter-Neutrino Resonance beginning at $r\sim 40\;{\rm km}$ and finishing at $r\sim 70\;{\rm km}$. As a result of the MNR, neither the neutrinos nor antineutrinos exhibit bipolar/nutation like transformation at $r \sim 150\;{\rm km}$ seen in the previous two examples. Thus we learn the MNR - which only occurs when we include NSI - stabilizes the system preventing collective effects. Once again the I and H resonance locations are predicted by the theory and we also observe how the numerical results track the expected evolution of the transition probabilities for neutrinos and antineutrinos from Eq.~(\ref{eq:MNRPred}) in the MNR region. 

\begin{figure}[t]
\includegraphics[width=0.47\textwidth]{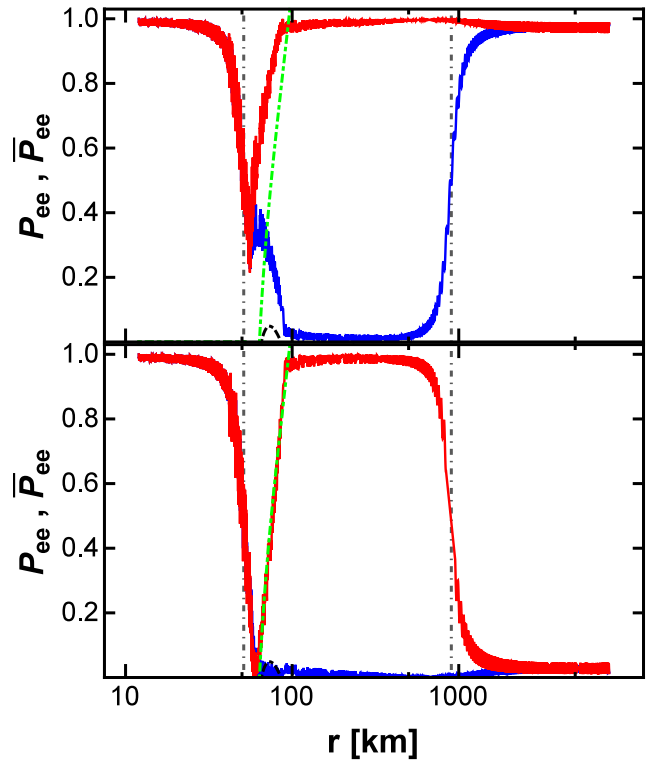}
\caption{The same as in figure (\ref{fig:MNRCase}) but with NSI parameters set to $\delta\epsilon^n = -0.9436$ and $\epsilon_0 = 0.0045$, corresponding to point C in figure~(\ref{fig:ParamOverview}). Here we see the effects of the increasing width of the I resonance causing an overlap with the Standard MNR, and a change in the behavior during the resonance.}
\label{fig:?MNRCase}
\end{figure}

\begin{figure}[t]
\includegraphics[width=0.47\textwidth]{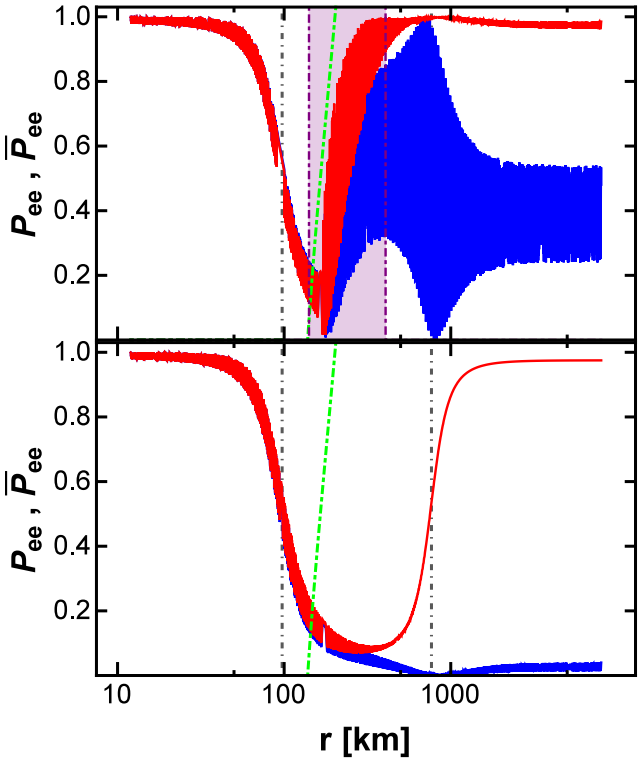}
\caption{The same as in figure (\ref{fig:MNRCase}) but with NSI parameters set to $\delta\epsilon^n = - 1.2124$ and $\epsilon_0 = 0.008$, corresponding to point D in figure~(\ref{fig:ParamOverview}). The vertical dot-dashed line at $r\approx 100\;{\rm km}= 10^{7}\;{\rm cm}$ is the I resonance and the vertical dot-dashed line at $r\approx 900\;{\rm km}$ is the H resonance. Here we see the I resonance followed by a nutation/bipolar transition. In this case, the width of the I resonance has completely covered the MNR suppressing it and preventing the system from stabilizing against the nutation region at $r\sim 150\:{\rm km}$.}
\label{fig:Bipolar2Case}
\end{figure}

The next set of NSI parameters we consider are $\delta\epsilon^n = -0.9436$ $\epsilon_0 = 0.0045$,corresponding to point C in figure~(\ref{fig:ParamOverview}), and the survival probabilities as a function of distance are shown in figure (\ref{fig:?MNRCase}). At this more negative value of $\delta\epsilon^n$ the I resonance has moved even further out to $r_I = 50\;{\rm km}$ and has grown noticeably wider so that it begins to overlap and interfere with the MNR. In the normal hierarchy (top panel) the I resonance only partially completes before the MNR begins; however in the inverted hierarchy the I resonance is allowed to complete fully before the MNR transition begins causing the MNR to narrow. We will discuss the difference of the behavior of the MNR in this figure compared to figure~(\ref{fig:MNRCase}) below. The H resonance can also be seen to have moved further inward as a consequence of the NSI but, as in previous figures, it remains adiabatic with almost a complete swap of e and x flavors. The predicted locations of the I and H resonances remains in good agreement with the numerical results; the predicted beginning and end of the MNR evolution is also in good agreement with the numerical results for both hierarchies but the actual evolution during the MNR is only well reproduced for the inverted hierarchy. 

\begin{figure}[t]
\includegraphics[width=0.47\textwidth]{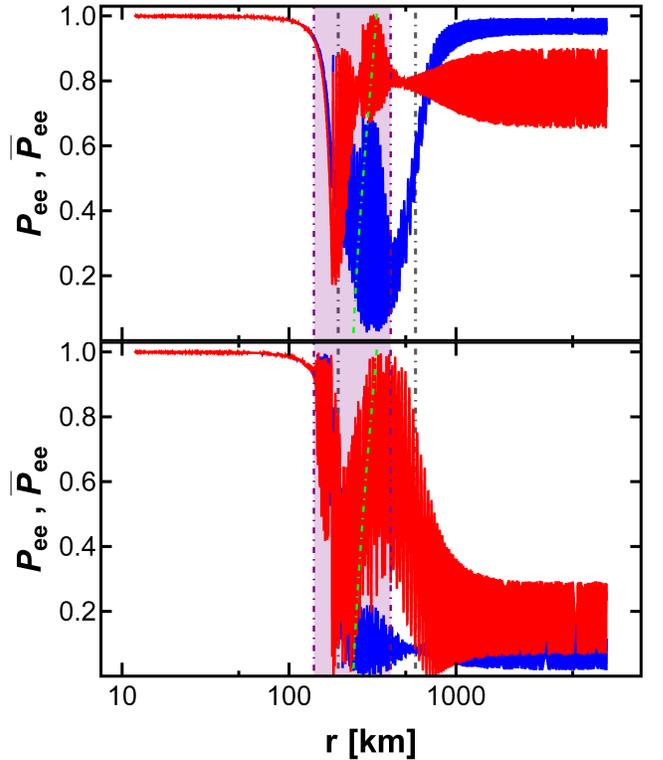}
\caption{The same as in figure (\ref{fig:MNRCase}) but with NSI parameters set to $\delta\epsilon^n = -1.4428$ and $\epsilon_0 = 0.00275$, corresponding to point E in figure~(\ref{fig:ParamOverview}). Here the resonance prediction lines show how tightly packed the various resonances have become causing a mixture of different behaviors the form of which changes with slight modifications to either $\delta\epsilon^n$ or $\epsilon_0$ in the yellow region.}
\label{fig:ChaoticCase}
\end{figure}

\begin{figure}[t]
\includegraphics[width=0.47\textwidth]{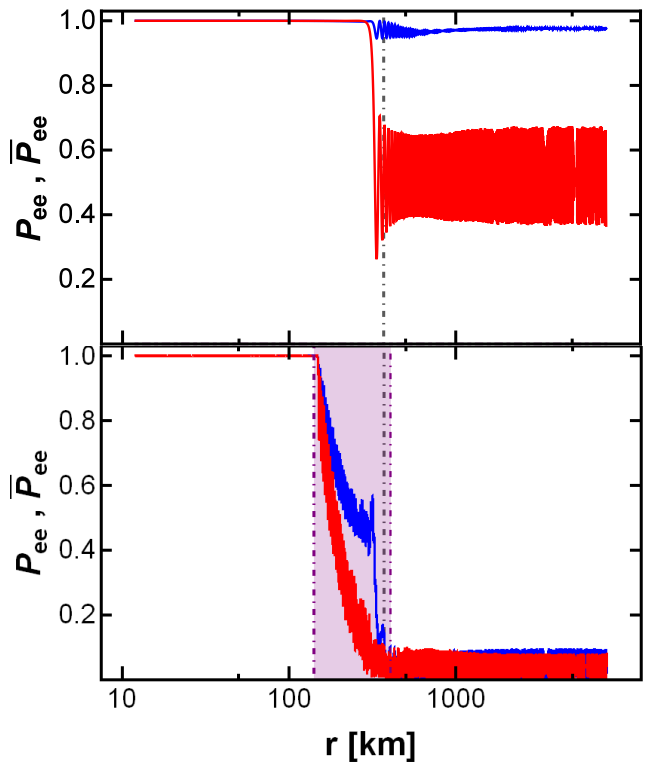}
\caption{The same as in figure(\ref{fig:MNRCase}) but with NSI parameters set to $\delta\epsilon^n = -1.6156$ and $\epsilon_0 = 0.0005$, corresponding to point F in figure~(\ref{fig:ParamOverview}). In this case there is no longer an I resonance or H resonance for the neutrinos in the normal hierarchy (nor or antineutrinos in the inverted hierarchy). Only the I resonance for the (anti)neutrinos in the normal (inverted) hierarchy remains along with the Bipolar/Nutation transition in the inverted hierarchy.}
\label{fig:NoMSWCase}
\end{figure}

In figure (\ref{fig:Bipolar2Case}) we plot the results for the NSI parameters $\delta\epsilon^n = -1.2124$ $\epsilon_0 = 0.008$, point D in figure~(\ref{fig:ParamOverview}). In this example the I resonance now occurs at $r_I\approx 100\;{\rm km}$ and has become noticeably wider than in previous examples. Beyond the I resonance we do not observe a Standard MNR but rather, in the normal hierarchy, a return to the bipolar/nutation behavior seen in figure (\ref{fig:Bipolar1Case}) and in the inverted hierarchy nothing happens until the H resonance.  Finally, we see in this example that the analytic predictions for the I and H resonances remain robust and, in addition to the outward motion of the I resonance, the simultaneous inward motion of the H resonance to $r_H \approx 900\;{\rm km}$. Both resonances remain adiabatic. 

We next consider point E in figure~(\ref{fig:ParamOverview}), with NSI parameters $\delta\epsilon^n = -1.4428$ and $\epsilon_0 = 0.00275$. The transition probabilities for this set of NSI parameters are shown in figure (\ref{fig:ChaoticCase}). Here we see all three types of transitions we have discussed, the I resonance, bipolar/nutation transition, and the MNR transition, are pushed very close together such that none can complete in a normal fashion. The figure includes the predictions from all three types of flavor transformation in an effort to help identify which features might belong to which effect. However it is very difficult to point to any clearly identifiable transition feature until we reach the H resonance at $r\sim 650\;{\rm km}$. 

Finally, in figure (\ref{fig:NoMSWCase}) we plot the survival probabilities of electron neutrinos and antineutrinos for $\delta\epsilon^n = -1.6156$ and $\epsilon_0 = 0.0005$, represented by point F in figure~(\ref{fig:ParamOverview}). For this combination of NSI parameters there is neither an I nor a H resonance for neutrinos in the normal hierarchy, and the same for antineutrinos in the inverted hierarchy. This is consistent with our analytical description as the resonance condition cannot be realized in this part of the parameter space. The remaining I resonance for antineutrinos in the normal hierarchy and neutrinos in the inverted hierarchy is pushed to large radii, beyond the start of the bipolar/nutation region. Since there is no I resonance before bipolar/nutation region, this instability region occurs in the inverted hierarchy, just as it does in the absence of NSIs. But not every effect of NSI has disappeared. Following the bipolar/nutation region we see the remaining I resonance at $r\sim 350\;{\rm km}$ but unlike in the standard oscillation case situation shown in figure (\ref{fig:NoNSI}), it acts upon the opposite state with the antineutrinos converting in the normal hierarchy and the neutrinos in the inverted hierarchy. 

\section{Partitioning the NSI parameter space}
\label{sec:partition}

\begin{figure*}
\includegraphics[width=\linewidth]{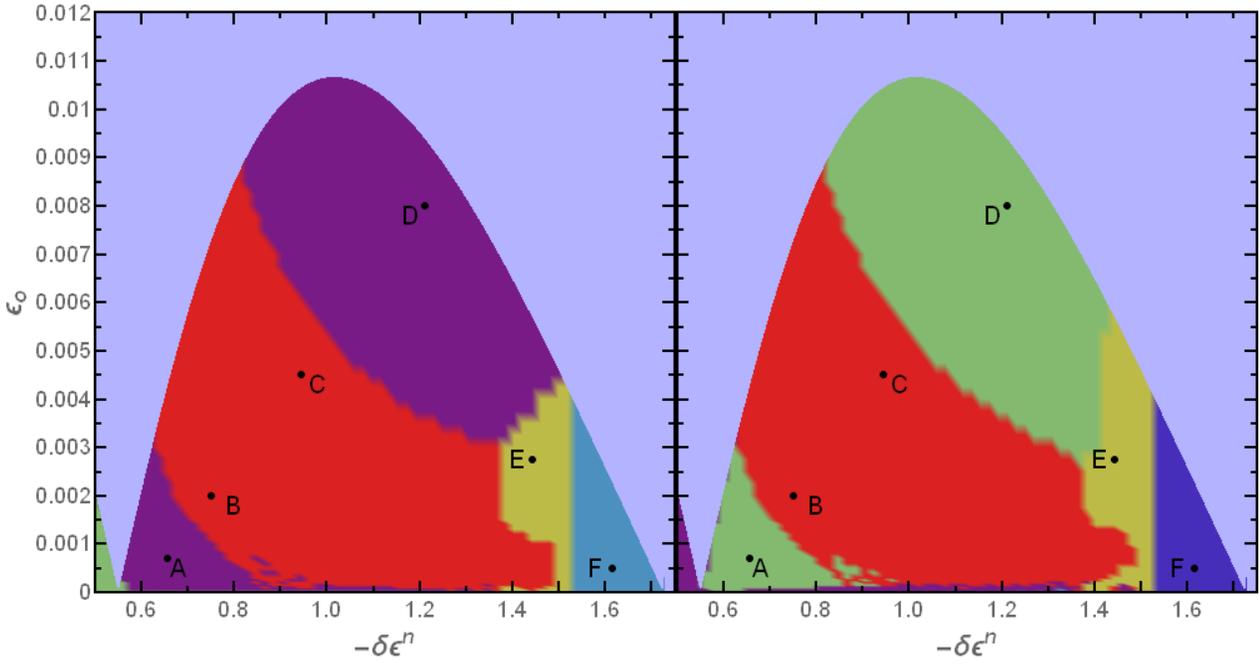}
\caption{The partition of the parameter space in the normal (right) and inverted (left) hierarchies. The different colored regions represent different behavior of the numerical solutions. The light blue region is the same as that shown in figure (\ref{fig:ParamOverview}), and is the region where the I resonance overlaps with the surface of the neutrinosphere. The Yellow Region shows where significant overlap between multiple resonances causes chaotic collective behavior. The soft blue region in the right panel shows where the I and H resonance disappear for neutrinos. The dark blue region in the left panel shows where the region where the I and H resonance disappear from antineutrinos after undergoing a bipolar transition. Green represents no collective oscillations between the I resonance and the H resonance. Purple represents results where we observed a bipolar/nutation transition. The Red regions are where MNR behavior is observed. The six black points represent the locations in parameter space of the example cases also seen in figure  (\ref{fig:ParamOverview}) whose survival probability plots are shown in figures (\ref{fig:Bipolar1Case}), (\ref{fig:MNRCase}), (\ref{fig:?MNRCase}), (\ref{fig:Bipolar2Case}), (\ref{fig:ChaoticCase}, and (\ref{fig:NoMSWCase}).}
\label{fig:ParameterPlots}
\end{figure*}

The various transformation effects seen in figures~(\ref{fig:Bipolar1Case}) - (\ref{fig:NoMSWCase}) are representative of behaviors seen through a wide range of NSI parameters in the magenta region shown in figure~(\ref{fig:ParamOverview}). The six figures shown in the previous section were taken from a larger scan of several thousand numerical calculations completed over this space. For each numerical run the results were compared to the analytical predictions described in \S \ref{sec:Matter}, as was done in the preceding section, and each resonance effect was identified. Figure~(\ref{fig:ParameterPlots}) was constructed by color coding each of the several thousand runs according to the resonance(s) that were observed. We will now use the same analytical tools to understand the shapes of the different resonance regions and the contours that naturally occur between them.

For $-1.52< \delta\epsilon^n < -0.55$ we observe in every calculation two I resonances - one in the neutrinos, the other in the antineutrinos - and an H resonance. In the inverted hierarchy we find two different regions where we see no other transitions occur within this range of $\delta\epsilon^n$, one at low values of $-\delta\epsilon^n$ and $\epsilon_0$, and the other at large values of both $-\delta\epsilon^n$ and $\epsilon_0$. These regions are shown in green in figure~(\ref{fig:ParameterPlots}). For this same region of the parameter space in the normal hierarchy we find bipolar/nutation type transitions also occur and these regions are shown in figure~(\ref{fig:ParameterPlots}) in purple. Over large swaths of the parameter space we observe some form of the MNR transition and these are indicated in the figure~(\ref{fig:ParameterPlots}) by the red regions. For $-1.52 < \delta\epsilon^n < -1.36$ we see chaotic effects where it can be difficult to determine which resonance transitions are playing a role in the final solution. This region is shown in yellow in figure~(\ref{fig:ParameterPlots}) and an example of such behavior was seen in figure~(\ref{fig:ChaoticCase}). For $\delta\epsilon^n < -1.52$ the flavor evolution simplifies greatly. Depending on the hierarchy the I resonance for either the neutrinos or antineutrinos is absent and the H resonance has disappeared as well. For the normal hierarchy there are no other types of flavor transformation; in the inverted hierarchy a bipolar/nutation type transition is seen which are sometimes interrupted by the remaining I resonance. The region of the NSI parameter space where this behavior occurs is depicted as the softer blue region in the right panel and the darker blue region in the left panel of figure~(\ref{fig:ParameterPlots}). 

\begin{figure}
\includegraphics{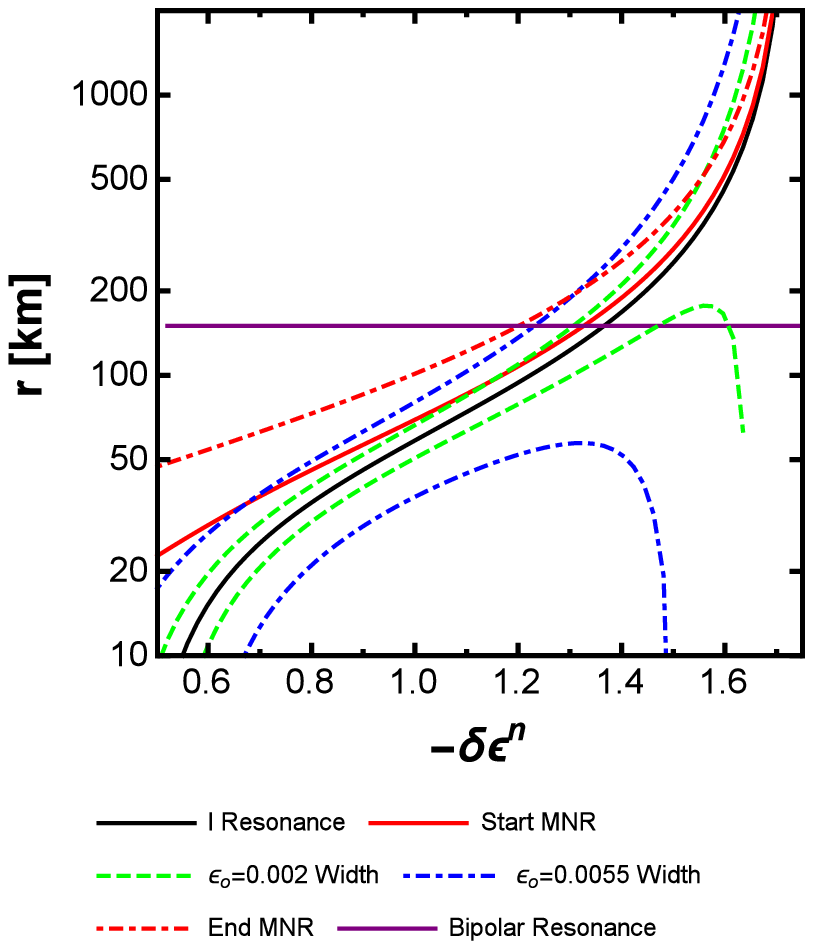}
\caption{The location of the I resonance, $r_I$ (solid black line), it's width $r_I \pm \sigma_I$ for two different values of $\epsilon_0$ (green dashed line, blue dot-dashed line), and the starting location (solid red line) and ending location (dot-dashed red line) of the MNR as a function of the NSI parameter $-\delta\epsilon^n$.}
\label{fig:ResPos}
\end{figure}

Clearly the effects of NSI are non-trivial with an interesting interplay between one kind of transformation and another. The analytic understanding of the various transformations from section \S\ref{sec:Matter} allow us to understand why various combinations of the NSI parameters give the transformations seen within each region of the parameter space. Let us begin with the leftmost purple and leftmost green regions from the two panels in figure (\ref{fig:ParameterPlots}). In this region the I resonance occurs close to the proto-neutron star and its effect is to swap the e and x spectra for both neutrinos and antineutrinos. After the I resonance the neutrinos could either undergo a Standard MNR or, for the case of the normal hierarchy, they could follow the bipolar/nutation transition because the swapping of the spectra has destabilized the neutrino system. In this region of the parameter space the Standard MNR condition is actually satisfied before the bipolar/nutation transition at $r=150 \;{\rm km}$ - for the representative point shown in figure (\ref{fig:Bipolar1Case}) the Standard MNR condition is satisfied at $r\approx 35\;{\rm km}$. But the Standard MNR transition only occurs if the neutrinos can evolve adiabatically and a glance at figure~(\ref{fig:VNSIvR}) indicates the gradients of both the total matter potential and neutrino-neutrino potential are very large at the location where the Standard MNR would begin. In this lower corner of the parameter space $\epsilon_0$ is not sufficiently large to allow a MNR transition to occur and if the MNR does not occur, then the bipolar/nutation transition takes place for the normal hierarchy. 

If $-\delta\epsilon^n$ and/or $\epsilon_0$ are increased, then  the neutrino evolution at the point where the Standard MNR condition is satisfied becomes more adiabatic. The increasing adiabaticity is visible in figure (\ref{fig:VNSIvR}) because one sees the increase of $-\delta\epsilon^n$ pushes the point where the MNR begins to larger radii, softening the gradients. Larger $\epsilon_0$ also allows for the MNR to occur with larger gradients. At sufficiently large $-\delta\epsilon^n$ and/or $\epsilon_0$ the Standard MNR occurs and the combinations of the NSI which lead to MNR transitions are the red regions in both the left and right panels of figure~(\ref{fig:ParameterPlots}). Through careful analysis of our results we find there is also a possibility for a Symmetric MNR to occur before the I resonance. As discussed in \S\ref{sec:Matter}, the Symmetric MNR can only occur in the region between the neutrinosphere and the location of the I resonance where a positive neutrino-neutrino potential and a negative total matter potential can cancel. While this transition is possible in principle, in the context of the model considered here we find minimal Symmetric MNR effects.

While an MNR is seen for all combinations of $\delta\epsilon^n$ and $\epsilon_0$ within the red bands, as both $-\delta\epsilon^n$ and $\epsilon_0$ are increased the MNR becomes less ideal. The difference can be seen by comparing figures~(\ref{fig:MNRCase}) and (\ref{fig:?MNRCase}) which are the results for the NSI parameters at the two points within the red bands in figure (\ref{fig:ParameterPlots}). This departure from ideal behavior is due to two different factors. As $-\delta\epsilon^n$ increases both the location of the I resonance and the beginning of the MNR, $r_{MNR,start}$ move outward, but they do so at different rates becoming ever closer. At the same time, an increase in $\epsilon_0$ causes the width of the I resonance, $\sigma_I$, to increase. Both of these trends can be observed in figure~(\ref{fig:ResPos}) which shows how the location and amount of overlap between the I resonance and the MNR changes with different values of $\delta \epsilon^n$ and $\epsilon_0$. At sufficiently large $-\delta\epsilon^n$ and $\epsilon_0$ these two factors will work together to cause the I resonance to partially overlap with the point at which the MNR is predicted to begin i.e.\ we find $r_I + \sigma_I > r_{MNR,start}$. When this condition is satisfied we observe the neutrinos and antineutrinos no longer follow the analytic expectations in Eq.~\eqref{eq:MNRPred} which assumed that both neutrinos and antineutrinos have fully converted with respect to their initial state before the start of the MNR. We note, as shown in figure~(\ref{fig:?MNRCase}), a difference in the effect of this overlapping of resonances between the normal and inverted hierarchies. In the normal hierarchy, the MNR appears to dominate over the I resonance, preventing it from completing; in the inverted hierarchy the I resonance appears to dominate over the MNR fully converting the neutrinos before the MNR begins. 

As both $-\delta\epsilon^n$ and $\epsilon_0$ are increased further the separation between the I resonance and MNR will continue to decrease and the width of the I resonance will continue to increase. Eventually a combination of NSI parameters will be reached such that the I resonance completely covers the MNR region. If $r_{MNR,end}$ is the predicted end point of the Standard MNR then the I resonance overwhelms the MNR when $r_I + \sigma_I > r_{MNR,end}$. The smothering of the MNR by the I resonance means no MNR occurs. For the normal hierarchy the stabilizing effect of the MNR is lost so the system will undergo a bipolar/nutation transition, for the inverted hierarchy no bipolar/nutaiton occurs because the system is stable. An example of this can be seen in figure~(\ref{fig:Bipolar2Case}) and to show how the I resonance now overlaps the region where the MNR occurs we have inserted the predicted MNR evolution into the figure. Thus at large $-\delta\epsilon^n$ and $\epsilon_0$ we again find bipolar/nutation transitions and this is the reason for the the upper purple region in the left panel of figure~(\ref{fig:ParameterPlots}). For the inverted hierarchy the combinations of $-\delta\epsilon^n$ and $\epsilon_0$ where the I resonance smothers the MNR are in the upper green region in the right panel of figure (\ref{fig:ParameterPlots}). 

In the light yellow shaded regions seen in figure (\ref{fig:ParameterPlots}) we find the chaotic evolution. This occurs when various transformation effects begin to overlap. In particular the I resonance moves inside of the instability region predicted by Eq.~\eqref{eq:stabcond}. From determining $r_I$ as a function of the NSI parameters we find $r_I = r_{bipolar,start}$ when $\delta\epsilon^n = -1.3658$. An example of this can be seen in figure~(\ref{fig:ChaoticCase}), where different prediction lines have been placed on the figure to help suggest which of the possible resonances might be the cause of the different features observed. 

Finally we consider the regions where $-\delta\epsilon^n > 1.52$ indicated by the light (dark) blue regions in the left (right) panel of figure~(\ref{fig:ParameterPlots}). At these values of $-\delta\epsilon^n$ the maximum value of the diagonal element of $V_M$ is less than the splitting of the vacuum eigenvalues and so there is no solution to Eq.~\eqref{eq:rI} for a normal hierarchy. Since the same equation must be fulfilled for the H resonance, the loss of the I resonance means the H resonance also disappears. The I resonance remains for the antineutrinos in the normal hierarchy because for any value of $\delta\epsilon^n$ the potential $\bar{V}_M$ has the same sign as the splitting of the vacuum eigenvalues and so the monotonicity of $\lambda(r)$ guarantees there must be a point where the two sides of the equation become equal. If the hierarchy is inverted, the I resonance only occurs for neutrinos and the I and H resonances disappear for antineutrinos when $-\delta\epsilon^n > 1.52$. The consequence of the missing I resonance is that only the neutrino spectra or only the antineutrino spectra are swapped, not both. Without swapping both spectra, the conditions for a Standard MNR cannot be fulfilled and for a normal hierarchy the system is stable. In the inverted hierarchy the system is unstable and with the model we are using in this paper we find that for $-\delta\epsilon^n > 1.52$ the remaining I resonance occurs after the bipolar/nutation begins at $r=150\;{\rm km}$. Thus in the inverted hierarchy the bipolar/nutation transformation begins but when the remaining I occurs, it shuts off the transition if it is not yet complete as can be seen in figure~(\ref{fig:NoMSWCase}). 

\begin{figure*}
\includegraphics[width=\textwidth]{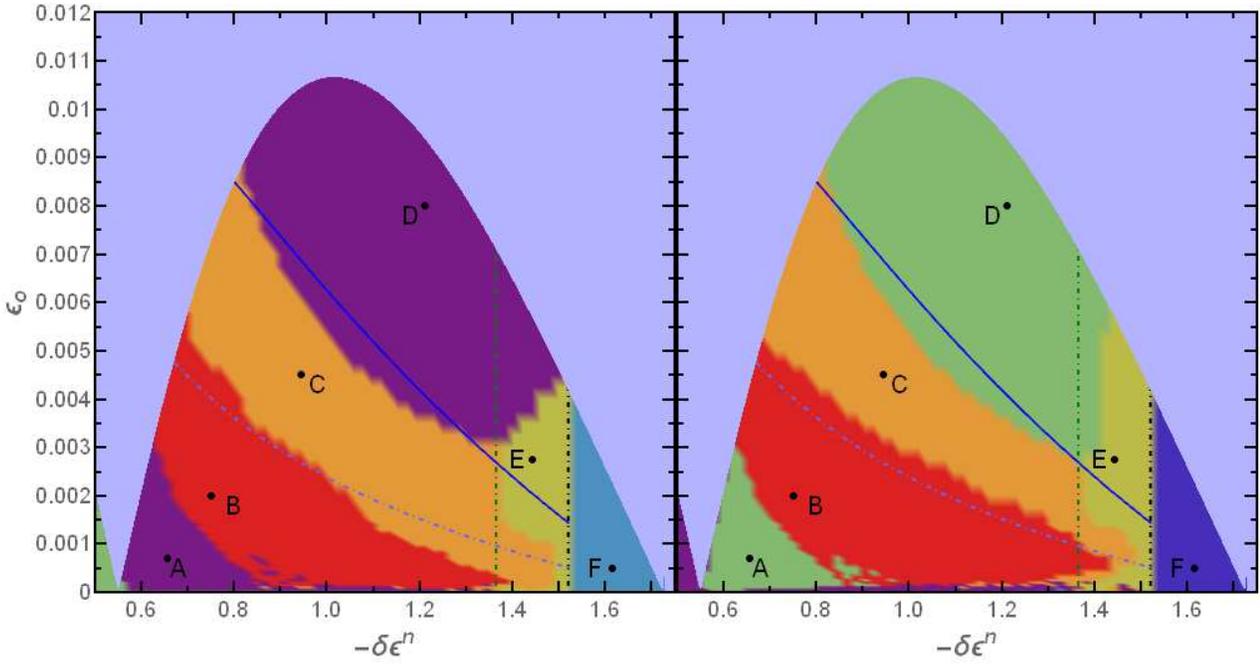}
\caption{Similar to figure (\ref{fig:ParameterPlots}) but with a further division of the MNR (red) region from figure~(\ref{fig:ParameterPlots}) into two regions. The red region now shows the parameter space where we observed the normal, complete, MNR as seen in figure~\ref{fig:MNRCase} and the orange region where the I resonance and MNR interfere with each other, as seen in figure~\ref{fig:?MNRCase}. The two lines roughly predicting the boundary between the full MNR and partial MNR (Dot-Dashed Blue) and where the solution transitions from MNR's back to bipolar transitions (Solid Blue). The green dot-dashed line represents the point that the predicted location of the I resonance lies within the instability region predicted by the stability matrix in Eq.~\eqref{eq:stab22}. The Black dot-dashed line is the value of $\delta\epsilon^n$ where the diagonal component of the matter potential no longer crosses the positive vacuum scale causing the I resonance and H resonance to disappear from the neutrinos, however the antineutrinos still undergo an I resonance. The other colors are the same as in figure (\ref{fig:ParameterPlots}): the light blue region is where the I resonance overlaps with the surface of the neutrinosphere, yellow is chaotic collective effects, soft blue is the absence of I and H resonances in normal hierarchy neutrinos, darker blue is the onset of the bipolar transition followed by an I resonance with no H resonance, green represents no collective oscillations between the I resonance and the H resonance and purple represents results where we observed a bipolar/nutation transition.}
\label{fig:FullParam}
\end{figure*}
Our understanding of the NSI effects allows us to place several lines upon the parameter space corresponding to where various resonances interact and thus where we expect to observe a given type of flavor transformation to become modified. These predictions and the partitioned parameter space are shown in figure~(\ref{fig:FullParam}). The various lines in the figure are 
\begin{itemize}
\item the contour where $r_I + \sigma_I = r_{MNR,start}$ (dashed blue)
\item the contour where $r_I + \sigma_I = r_{MNR,end}$ (solid blue)
\item the value of $\delta\epsilon^n=-1.32$ which gives $r_I = r_{bipolar,start}$ (dashed green)
\item and the vertical line $\delta\epsilon^n=-1.521$ where the H resonance disappears. (dashed black)
\end{itemize}
In addition to these predictions, we have separated the red band from figure~(\ref{fig:ParameterPlots}) into a region again shaded red containing the ideal, `complete' MNR solutions, and an orange region where significant overlap between the I resonance and the MNR is found leading to `partial' MNRs. It should be noted that the transition from the complete MNR to the partial MNR is not as sharp as figure~(\ref{fig:FullParam}) makes it appear. Rather there is a much more gradual change from cases such as figure~(\ref{fig:MNRCase}) to the significantly overlapped case seen in figure~(\ref{fig:?MNRCase}). 

The first contour is our prediction for the combinations of $\delta\epsilon^n$ and $\epsilon_0$ where we expect the I resonance to \emph{begin} to overlap with the MNR and so lead to partial MNR like solutions, e.g. figure~(\ref{fig:?MNRCase}). This contour is shown as the dot-dashed blue line. The second contour identifies the values of $\delta\epsilon^n$ and $\epsilon_0$ where the I resonance \emph{completely} overlaps the MNR at which point the system should transition back to its behavior in the absence of the MNR i.e.\ a bipolar/nutation transition for a normal hierarchy and a no collective oscillations for the inverted. This contour is shown as the solid blue line. The two contours trace the boundaries between the different categories of solutions well. The third condition is the location where the I resonance and beginning of the bipolar/nutation regions coincide. It is the merging of the different resonances that creates the somewhat chaotic flavor evolution observed in figure~(\ref{fig:ChaoticCase}).The figure shows how this line delimits one edge of the yellow region. The fourth condition denotes where the H resonance disappears and this vertical line matches well with the edge of the blue regions. 

\section{Discussion and Conclusions}
\label{sec:conc}

In this paper we have shown how Nonstandard Interactions of neutrinos, well within current constraints, can lead to dramatically different flavor evolution for supernova neutrinos compared to both the standard case of neutrino oscillations and previous literature. In a broad region of the parameter space we observe Matter-Neutrino Resonances for supernova neutrinos, a type of flavor transformation which has previously only been seen in compact object merger scenarios. In another region of the parameter space we find the NSI can lead to neutrino collective effects which would not occur in the absence of the NSI and can suppress other collective effects which would be expected using standard neutrino oscillation physics. Finally, in a third region we find the NSI can lead to the disappearance of the high density Mikheyev-Smirnov-Wolfenstein resonance. From our understanding of how these effects arise we are able to predict the boundaries between the partitions of the NSI parameter space where the various transformations are seen. 

Such dramatic flavor transformation due to NSI so deep within the supernova has the potential to affect the dynamics of the explosion, the nucleosynthesis and the neutrino burst signal. If only the I resonance occurs one can make plausible predictions for the effect of the NSI because, by itself, an adiabatic I resonance leads to a complete swap of the flavor of both neutrinos and antineutrinos. Beyond the I resonance the spectrum of the electron flavor neutrinos and antineutrinos would be hotter than that at the neutrinosphere and one would expect this would lead to greater heating in the gain layer and a shorter delay until shock revival. The additional flavor transformation effects which occur, such as the MNR and bipolar/nutation, modify this expectation. If the neutrinos undergo a MNR transition then the antineutrino flavors swap back to their original spectra while the neutrinos remain swapped. During a bipolar/nutation transformation both neutrino and antineutrinos re-exchange the spectra but not as completely as during an MNR. If the I resonance is not adiabatic then an MNR cannot occur but a bipolar/nutation can. An understanding of the dependence of supernova dynamics upon the NSI parameters will require further study.

Similarly, flavor transformation so deep within a supernova will affect the electron fraction of the material and the subsequent nucleosynthesis. If the I resonance is adiabatic the neutrinos and antineutrinos both swap the spectra of their flavors. As a consequence the spectra of both the electron neutrinos and antineutrinos will be more similar than the original spectra were at the neutrinosphere which should raise the electron fraction slightly compared to the unoscillated case. But if the I resonance is quickly followed by a MNR then, again, the antineutrino flavors can swap back to their original spectra while the neutrinos remain altered. Examples of these cases are shown in figures (\ref{fig:MNRCase}) and (\ref{fig:?MNRCase}). The difference between the electron neutrino and electron antineutrino spectra would then be larger than at the neutrinosphere. Supernova wind nucleosynthesis is sensitive to the difference between the electron neutrino and electron antineutrino spectra so the question of how the nucleosynthesis in supernovae might be modified by NSI also needs to be addressed in future studies.

Finally, the effects of NSI clearly alter the flux emerging from the supernova and the conclusions one might draw from the next Galactic supernova burst signal. Features in the signal which are associated with one hierarchy in the standard case of neutrino oscillations can instead occur in the other hierarchy when NSI are included and move from neutrino to the antineutrino channels. The disappearance of the H resonance that can occur in some regions of the NSI parameter space will also have profound effects upon the signal. The signatures of NSI effects in a Galactic supernova neutrino burst and the detector requirements to observe them will need to be determined.

\appendix*
\section{Analytical MNR Survival Probabilities}
\label{sec:appendix}

The analytical expressions in Eq.~\eqref{eq:MNRPred} for the neutrino and antineutrino survival probabilities during an MNR are derived from the MNR conditions ~\cite{2015arXiv151000751V}. First we demand $0 = H_{ee} - H_{xx}$ and using the definitions for the vacuum Hamiltonian, the neutrino-neutrino interaction potential, and the matter potential, we write this condition as
\begin{equation}
\begin{aligned}
0 =& \frac{\delta m^2}{2E} \cos2\theta_V + \lambda Y_e + \lambda\delta\epsilon^n\left(\frac{Y_{\odot}-Y_e}{Y_{\odot}}\right) \\ 
& + \mu_{\nu}\,\left((1+\beta)(\rho_{ee}-\rho_{xx}) - \alpha (1+\bar{\beta})(\bar{\rho}_{ee} - \bar{\rho}_{xx})\right).
\label{eq:MNRCOND}
\end{aligned}
\end{equation}
We have suppressed the spatial dependence of $\lambda$, $Y_e$, $\mu_{\nu}$, and the density matrix elements for clarity. The second MNR condition is that the off-diagonal components of the Hamiltonian must vanish ~\cite{2015arXiv151000751V} which leads to 
\begin{eqnarray}
0& = &\frac{\delta m^2}{2E} \sin2\theta_V + \lambda (3+Y_e)\epsilon_0\nonumber \\ 
&&+ \mu_{\nu}\left((1+\beta)\rho_{ex} - \alpha(1+\bar{\beta})\bar{\rho}_{xe}\right).
\label{eq:OffDiag}
\end{eqnarray}
These two equations can be used to derive how the survival probabilities of the neutrinos and antineutrinos must change in order to continue to satisfy the two conditions of the MNR.

These equations can be greatly simplified using the common assumptions in MNR scenarios. First we assume that the vacuum term is negligible compared to $V_M$ or $V_{\nu}$ and, second, we assume that $\lambda(r)\epsilon_0 \ll \mu_{\nu}(r)$ allowing us to further simplify the condition on the off-diagonal components of the Hamiltonian. Using these assumptions, we can rewrite Eq.~\eqref{eq:MNRCOND} and Eq.~\eqref{eq:OffDiag} as
\begin{eqnarray}
\label{eq:EQA}
0 & = & q+\left((1+\beta)(\rho_{ee}-\rho_{xx}) - \alpha(1+\bar{\beta})(\bar{\rho}_{ee} - \bar{\rho}_{xx})\right)  \nonumber \\ & & \\
0 & = & \left((1+\beta)\rho_{ex} - \alpha(1+\bar{\beta})\bar{\rho}_{xe}\right)
\label{eq:EQB}
\end{eqnarray}
where 
\begin{equation}
q = \frac{V_{M,ee}}{\mu_{\nu}} = \frac{\lambda}{\mu_{\nu}}\left(Y_e + \delta\epsilon^n\left(\frac{Y_{\odot} - Y_e}{Y_{\odot}}\right)\right) .
\end{equation}
This is the same definition of $q$ found in Eq.~\eqref{eq:MNRPred}. 

The most efficient path to solving for the survival probabilities makes use of isospin vectors. In order to make the connection between isospin vectors and the density matrices we factor the combined density matrix, $\rho$ used throughout the body of this paper into two density matrices which are initially pure flavor states, $\rho^e$ and $\rho^x$. More precisely we write
\begin{eqnarray}
\label{eq:exDef}
\rho(r) &=& \frac{1}{1+\beta}\left(\rho^e(r) + \beta \rho^x(r)\right)\\ 
\rho^e(0) &=& \begin{pmatrix}1 & 0 \\ 0 & 0\end{pmatrix}\\
\rho^x(0) &=& \begin{pmatrix}0&0\\0&1\end{pmatrix}
\label{eq:SeperateIC}
\end{eqnarray}
with similar expressions for $\bar{\rho}$, $\bar{\rho}^e$, and $\bar{\rho}^x$. The density matrices $\rho^e$ and $\rho^x$ are related to isospin vectors for the e and x flavor neutrinos, $\vec{S}_e$ and $\vec{S}_x$ respectively (with $\vec{\bar{S}}_e$ and $\vec{\bar{S}}_x$ for antineutrinos) defined to be
\begin{equation}
\vec{S}_n = \begin{pmatrix}\Re(\rho^n_{ex}) \\ -\Im (\rho^n_{ex}) \\ \frac{1}{2}(\rho^n_{ee} - \rho^n_{xx}). \end{pmatrix}
\end{equation}
This definition ensures $\vec{S}_e$ and $\vec{S}_x$ follow the normal rules for the isospin vector i.e.~ $S_n^2 = \vec{S}_n\cdot\vec{S}_n = \frac{1}{4}$.
Using $\vec{S}_e$ and $\vec{S}_x$ we can write the isospin vector $\vec{S}$ - defined using the combined density matrix $\rho$ and which is not normalized to $\vec{S}\cdot\vec{S}=\frac{1}{4}$ - in terms of $\vec{S}_e$ and $\vec{S}_x$ and find the product $\vec{S}\cdot\vec{S}$ is equal to
\begin{equation}
\begin{aligned}
S^2 =& \left(\vec{S}_e + \beta \vec{S}_x\right)^2 = \vec{S}_e^2 + 2\beta\vec{S}_e\cdot\vec{S}_x + \beta^2\vec{S}_x^2\\
=& \left[\left(\Re(\rho^e_{ex})+\beta \Re(\rho^x_{ex})\right)^2 + \left(\Im(\rho^e_{ex})+\beta \Im(\rho^x_{ex})\right)^2\right] \\ 
 &+ \frac{1}{4}\left(\rho^e_{ee} + \beta\rho^x_{ee} - \rho^e_{xx} - \beta\rho^x_{xx}\right)^2
\label{eq:S1}
\end{aligned}
\end{equation}
We simplify Eq.~\eqref{eq:S1} by recombining elements of $\rho^e$ and $\rho^x$ into the elements of $\rho$ according to Eq.~\eqref{eq:exDef}, apply the invariance of $S_e^2$ and $S_x^2$ and use the initial conditions for $\rho^e$ and $\rho^x$. After this simplification we find Eq.~\eqref{eq:S1} becomes
\begin{eqnarray}
(1-\beta)^2 & = & 4\left[(1+\beta)\Re(\rho_{ex})\right]^2 + 4\left[(1+\beta)\Im(\rho_{ex})\right]^2\nonumber \\ 
&&+ \left[(1+\beta)(\rho_{ee} - \rho_{xx})\right]^2.
\label{eq:SDef}
\end{eqnarray}
An identical procedure can be followed for the antineutrinos to obtain
\begin{eqnarray}
(1-\bar{\beta})^2 & = & 4\left[(1+\bar{\beta})\Re(\bar{\rho}_{ex})\right]^2 + 4\left[(1+\bar{\beta})\Im(\bar{\rho}_{ex})\right]^2 \nonumber \\ 
&& + \left[(1+\bar{\beta})(\bar{\rho}_{ee} - \bar{\rho}_{xx})\right]^2
\label{eq:SbarDef}
\end{eqnarray}
To make further progress we introduce the following definitions:
\begin{equation}
\begin{aligned}
(1+\beta)\rho_{ee} &= Z_e\\
(1+\beta)\rho_{xx} &= (1+\beta)(1-\rho_{ee}) = 1+\beta-Z_e\\
(1+\beta)\rho_{ex} &= \delta + i \epsilon\\
(1+\bar{\beta})\bar{\rho}_{ee} &= X_e\\
(1+\bar{\beta})\bar{\rho}_{xx} &= (1+\bar{\beta})(1-\bar{\rho}_{ee}) = 1+\bar{\beta}-X_e\\
(1+\bar{\beta})\bar{\rho}_{ex} &= \eta + i \theta
\end{aligned}
\end{equation}

Using these definitions we find Eqs.~\eqref{eq:EQA}, \eqref{eq:SDef}, \eqref{eq:SbarDef}, and \eqref{eq:EQB} reduce to the following system of equations
\begin{equation}
\begin{aligned}
0 & = q+2Z_e-(1+\beta)-\alpha\left(2X_e-(1+\bar{\beta})\right)\\
(1-\beta)^2 & = 4\left[\delta^2+\epsilon^2\right]+(2Z_e-(1+\beta))\\
(1-\bar{\beta})^2 & = 4\left[\eta^2 + \theta^2\right]+(2X_e-(1+\bar{\beta}))\\
0 & = \delta-\alpha\eta\\
0 & = \epsilon + \alpha \theta\\
\end{aligned}
\end{equation}
where we satisfy the real and imaginary parts independently. The last two of these equations can be combined to create an equality between the second and third equations, giving us two equations that involve only $X_e$ and $Z_e$. When we solve these equations we find
\begin{equation}
\begin{aligned}
Z_e &= \frac{1+\beta}{2} + \frac{\alpha^2(1-\bar{\beta})^2 - (1-\beta)^2-q^2}{4\,q}\\ 
X_e &= \frac{1+\bar{\beta}}{2} + \frac{\alpha^2(1-\bar{\beta})^2-(1-\beta)^2+q^2}{4\,\alpha\,q}
\end{aligned}
\end{equation}
Using the definition of $Z_e$ and $X_e$ as well as the equations for the survival probabilities given in Eq.~\eqref{eq:SurvivalProbabilities}, we can convert these results into the analytical expressions for the survival probabilities and thus derive Eq.~\eqref{eq:MNRPred}
\begin{equation}
\begin{aligned}
P_{ee} &= \frac{1}{2}\left(1+\frac{\alpha^2(1-\bar{\beta})^2 - (1-\beta)^2-q^2}{2\,(1-\beta)\,q}\right)\\
\bar{P}_{ee} &= \frac{1}{2}\left(1+\frac{\alpha^2(1-\bar{\beta})^2 - (1-\beta)^2+q^2}{2\,\alpha\,(1-\bar{\beta})\,q}\right)
\end{aligned}
\end{equation}
In the limit where $\beta \to 0$, $\bar{\beta}\to0$, and we set $V_{NSI}\to 0$ (which forces $q \to R$), we recover Eq.~\eqref{eq:OldMNR}. 

\begin{acknowledgments}
This research was supported by DOE Grants No. DE-SC0006417, No. DE-FG02-10ER41577, and No. DE-FG02-02ER41216, the Undergraduate Research in Computational Astrophysics Research Experience for Undergraduates program at North Carolina State funded by the NSF under No. AST-1032726, and  US Department of Education Graduate Assistance in Areas of National Need (GAANN) grant number P200A150035.
\end{acknowledgments}

\bibliographystyle{apsrev4-1}
\bibliography{NSI}

\end{document}